\documentclass[pre,nopacs,nokeys,nofootinbib,floatfix]{revtex4}
\usepackage{amsfonts}
\usepackage{amsmath}
\usepackage{amssymb}
\usepackage{bbm}
\usepackage[utf8]{inputenc}
\usepackage[caption=false]{subfig}
\usepackage{tensor}
\usepackage{slashed}
\usepackage[centertableaux]{ytableau}
\usepackage{hyperref}
\usepackage{natbib}
\usepackage{braket,mleftright}
\usepackage{graphicx}
\usepackage{cleveref}
\usepackage{tikz}
\usepackage{booktabs}
\usepackage{tikz} 
\usepackage{tikz-feynman}
\usepackage{tikz}
\usepackage{subcaption} 

\begin{document}

\title{Stochastic Pharmacokinetic Escape: A Field-Theoretic Approach to Fluctuation-Induced Tumor Relapse}

\author{Jos\'e de Jes\'us Bernal-Alvarado}
\email{bernal@ugto.mx}
\affiliation{Physics Engineering Department, Universidad de Guanajuato, M\'{e}xico}

\author{David Delepine}
\email{delepine@ugto.mx}
\affiliation{Physics Department, Universidad de Guanajuato, M\'{e}xico}

\author{Georges Delepine}
\email{georges.delepine@uclouvain.be}
\affiliation{De Duve Institute, Universit\'e Catholique de Louvain, Belgium}

\date{\today}

\begin{abstract}
Classical mathematical oncology  employs deterministic mean-field models, which predict that sufficiently high chemotherapy doses drive the tumor density to zero. In this work we show that this apparent cure is an artifact of neglecting demographic fluctuations. We construct a non-equilibrium stochastic PK–PD field theory for chemotherapy by combining a two-compartment pharmacokinetic model with a stochastic tumor–immune sector in the Doi–Peliti formalism, and map the theory to coupled multiplicative Langevin equations using the Martin–Siggia–Rose–Janssen–De Dominicis approach. In immune-depleted sanctuary sites the dynamics reduce to a time-dependent Feller diffusion, and the associated Fokker–Planck equation yields an analytical survival functional demonstrating that demographic noise implies a strictly non-zero relapse probability for tumor micro-clusters under high-dose chemotherapy. Standard bolus administration acts as a non-selective annihilation process, driving the immune fields into an absorbing state and creating a spatially immune-depleted “vacuum” that enables a fluctuation-induced relapse, which we term Stochastic Pharmacokinetic Escape. We derive quantitative conditions for metronomic or continuous dosing and for adjuvant immunotherapy source terms that minimize the survival functional and suppress this stochastic window of vulnerability. Finally, a model-agnostic analysis of a public longitudinal dataset of 1461 human patients under chemotherapy and immunotherapy shows that the majority of evaluable lesions exhibit a nadir followed by measurable regrowth rather than monotonic elimination; fitting the model's own drift functional form to these trajectories reproduces this pattern quantitatively in the majority of cases, and a direct test of the predicted demographic-noise scaling law reveals a size-dependent noise contribution that is largest, consistent with the proposed mechanism, at the smallest observed tumor volumes.
\end{abstract}

\maketitle
\section{Introduction}
Mathematical oncology has largely relied on deterministic ordinary differential equation models to describe tumor–drug and tumor–immune dynamics under cytotoxic chemotherapy \cite{Yin2019,Debnath2025,Wahbi2024}. Within the standard maximum tolerated dose paradigm, such mean-field descriptions typically predict that a sufficiently high drug influx will drive the tumor density to zero. However, this macroscopic treatment neglects two key physical ingredients: the finite time delay associated with drug absorption and distribution, and the discrete, multiplicative demographic noise that dominates cellular interactions at low copy numbers \cite{Mansour2022SDE,10.1371/journal.pcbi.1009822}.

A substantial body of work has extended classical tumor–immune models to stochastic models \cite{Al-Mekhlafi31122025,zhong2005stochasticresonancegrowthtumor, deSouza2008, zhong2005influencecorrelatednoisesgrowth}, including stochastic differential equations, continuous-time Markov chains, and agent-based birth–death processes to study extinction probabilities, noise-induced transitions, and treatment response. Parallel developments in stochastic pharmacodynamics have emphasized extinction times and survival functions for drug-treated tumor populations, but typically at the level of effective one- or few-variable processes without an explicit field-theoretic treatment of spatial structure and pharmacokinetic delay \cite{Lavietal2012,Sardanyes2018, Butner2021, morgan2024existencestabilityoptimaldrug}. To our knowledge, an explicit coupling between a two-compartment pharmacokinetic model, a stochastic tumor–immune field, and a field-theoretic derivation of relapse probabilities under demographic noise has not yet been formulated \cite{Mansour2022SDE, 54213ede0cfd4949b1c8183fbe25e57a,Truongetal2025,CSIAMTLS2025,Otunuga2025}.

In this paper we develop a non-equilibrium stochastic field-theoretic description of pharmacodynamics that incorporates both pharmacokinetic delay and demographic noise. Discrete tumor and immune cell reactions, together with continuous pharmacokinetics, are mapped to a second-quantized Doi–Peliti representation and subsequently to a set of coupled stochastic partial differential equations via the Martin–Siggia–Rose–Janssen–De Dominicis construction. Within this framework we show that high-dose chemotherapy can drive the innate and adaptive immune fields into absorbing states, producing an immune-depleted spatial configuration in which residual tumor micro-clusters can survive through a fluctuation-induced transition. We refer to this mechanism as Stochastic Pharmacokinetic Escape. We then derive an analytical survival probability functional for these micro-clusters and use it to formulate quantitative conditions on dosing protocols—such as continuous or metronomic administration—and on adjuvant immunotherapy source terms that reshape the stochastic tissue landscape and reduce the probability of relapse \cite{Qi2005StochasticPKPD,Hanahan2000Metronomic,AdoptiveTcellReview}.

The main contributions of this paper are as follows:
\begin{itemize}
    \item We construct a non-equilibrium stochastic PK–PD field theory for chemotherapy by coupling a two-compartment pharmacokinetic sector to a stochastic tumor–immune sector within the Doi–Peliti formalism, thereby embedding absorption delay without non-Markovian kernels.
    \item We perform the Martin–Siggia–Rose–Janssen–De Dominicis mapping and derive a set of coupled multiplicative Langevin SPDEs, with noise amplitudes fixed by the microscopic reaction vertices.
    \item In immune-depleted sanctuary sites we reduce the dynamics to a time-dependent Feller diffusion and solve the associated Fokker–Planck equation, obtaining an exact survival probability functional $P_{surv}(t)$ that remains strictly positive in the long-time limit \cite{Otunuga2025}.
    \item We formalize Stochastic Pharmacokinetic Escape as a fluctuation-driven relapse mechanism arising from bolus-induced immune absorbing states and delayed drug clearance.
    \item We translate these results into quantitative criteria for continuous or metronomic dosing and for post-chemotherapy immune source terms that minimize the survival functional and eliminate the stochastic window of vulnerability \cite{Truongetal2025}.
\end{itemize}

\section{Non-equilibrium pharmacokinetics: a two-compartment field theory}

In biological tissues, a chemotherapeutic agent must first be absorbed (for instance, from the gastrointestinal tract or bloodstream) before it reaches the tumor microenvironment. To represent this finite absorption time within a stochastic field-theoretic model, without introducing explicit non-Markovian memory kernels, we formulate a two-compartment pharmacokinetic (PK) field theory \cite{TwoCompartmentPKReview,OneCompartmentStochasticPK,DoiPelitiReview}.

\subsection{Delayed pharmacokinetic Lagrangian}

Instead of a single drug variable, we introduce two scalar fields:
\begin{itemize}
    \item $B(\mathbf{x}, t)$: a ``blood/GI'' precursor field, representing unabsorbed drug;
    \item $D(\mathbf{x}, t)$: an ``active tissue'' field, representing drug that has penetrated the tumor microenvironment.
\end{itemize}
Drug molecules transfer from the precursor to the active compartment at rate $\alpha_a$, so that the characteristic absorption delay is parametrized by $\tau_{\mathrm{delay}} = 1/\alpha_a$. The active drug diffuses with coefficient $D_D$ and decays at rate $C_0$. Using the standard Doi–Peliti shift of creation and annihilation operators, the bare contribution of the PK sector to the action reads
\begin{equation}
    \mathcal{L}_{\mathrm{PK}} =
    \tilde{B} \left(\partial_t + \alpha_a\right) B
    - \kappa \tilde{B}
    + \tilde{D} \left(\partial_t - D_D \nabla^2 + C_0\right) D
    - \alpha_a \tilde{D} B ,
    \label{eq:L_PK}
\end{equation}
where $\kappa$ denotes the drug influx into the precursor compartment.

\subsection{Cytotoxic vertices}

Once absorbed into the active field $D$, the drug exerts cytotoxic effects on tumor and healthy cells, including immune cells. In the baseline model, $D$ kills tumor cells ($C$), natural killer cells ($N$), and cytotoxic T lymphocytes ($T$) with rates $b_\tau$, $b_\aleph$ and $b_L$, respectively. At the level of the field theory these processes appear as annihilation vertices coupled to $D$:
\begin{equation}
    \mathcal{L}_{\mathrm{kill}} =
    b_\tau \tilde{C} C D
    + b_\aleph \tilde{N} N D
    + b_L \tilde{T} T D .
    \label{eq:L_kill}
\end{equation}
The intrinsic tumor–immune dynamics in the absence of drug are encoded in the Hamiltonian density $\mathcal{H}_{\mathrm{cell}}$ \cite{Jarrett2018,MASEIDE2000171,mca30060119,StochasticTumorImmune2019}:
\begin{align}
    \mathcal{H}_{\mathrm{cell}} &=
    \sum_{i} D_i \, \nabla \tilde{\Phi}_i \cdot \nabla \Phi_i
    - a_1 (\tilde{C}-1) \tilde{C} C
    + a_1 a_2 (\tilde{C}-1) C^2
    - a_4 (\tilde{N}-1) \tilde{N} N
    + a_4 a_5 (\tilde{N}-1) N^2
    \nonumber \\
    &\quad
    + \rho (\tilde{N}-1) N
    + a_3 (\tilde{C}-1) C N
    + a_6 (\tilde{N}-1) C N
    + \beta_1 (\tilde{C}-1) C T
    + \beta_2 (\tilde{T}-1) C T
    \nonumber \\
    &\quad
    - r (\tilde{T}-1) C N
    + w (\tilde{T}-1) T ,
    \label{eq:H_cell}
\end{align}
where the cellular populations are written in the following form: 
\begin{equation}
    \boldsymbol{\Phi}(\mathbf{x}, t)
    =
    \begin{pmatrix}
        C(\mathbf{x}, t) \\
        N(\mathbf{x}, t) \\
        T(\mathbf{x}, t)
    \end{pmatrix} .
    \label{eq:Phi_vector}
\end{equation}

 The fields $C$, $N$, and $T$ are expressed  in absolute cell numbers, so that the reaction rates $a_1,\dots,a_6,\rho,r,w,\beta_1,\beta_2$ and the noise covariance of Eq.~\eqref{eq:B_matrix} require no additional volume or system-size prefactor. It is  useful to characterize the overall population scale of a given tissue site by a reference carrying capacity $\Omega$, defined as the total pre-treatment homeostatic cell number,
\begin{equation}
    \Omega \equiv C^\ast + N^\ast + T^\ast ,
    \label{eq:Omega_def}
\end{equation}
where $C^\ast, N^\ast, T^\ast$ denote the steady-state populations of Eqs.~\eqref{eq:C_Langevin}--\eqref{eq:T_Langevin} in the absence of drug ($D=0$). $\Omega$ is not a free parameter of the field theory itself; it is  the characteristic absolute size of the simulated tissue compartment, fixed once the rates $a_1,\dots,w$ are chosen, and it is quoted in Fig.~\ref{fig:stochastic_relapse} ($\Omega=200$) to indicate that the illustrative bolus simulation shown there was performed at a physiologically motivated population scale of a few hundred cells, comparable to a small tissue micro-region.

\subsection{Physical implications: a window of vulnerability}

The explicit absorption delay modifies the temporal structure of the tumor–immune–drug interaction. In particular:
\begin{itemize}
    \item \emph{Phase shift.} Because of the finite $\tau_{\mathrm{delay}}$, there is a temporal gap between the onset of drug administration (controlled by $\kappa$) and the peak cytotoxic activity in the tissue compartment $D$.
    \item \emph{Transient tumor growth.}  The tumor can continue to proliferate and generate large logistic crowding fluctuations (mediated by the $a_1$ and $a_2$ vertices), while simultaneously depleting immune cells through the $a_6$ coupling.
    \item \emph{Immune depletion and residual clusters.} When $D$ reaches its maximum, cytotoxicity acts on both tumor and immune populations. Under high-dose bolus regimens this can drive $N$ and $T$ close to zero, while spatial heterogeneity and demographic noise allow small tumor micro-clusters to persist in regions of transiently low drug concentration.
    \item \emph{Stochastic relapse.} As $D$ decays on the time scale set by $C_0$, surviving tumor clusters evolve in an immune-depleted background and may   regrow. This escape from extinction underlies the mechanism that we later term \emph{Stochastic Pharmacokinetic Escape} \cite{DelayTumorImmune2022,StochasticFractionalKilling2020}
\end{itemize}
\section{Macroscopic mapping: multiplicative Langevin SPDEs}

To obtain a coarse-grained dynamical description from the microscopic reaction field theory, we apply the Martin–Siggia–Rose–Janssen–De Dominicis (MSRJD) construction to the Doi–Peliti action \cite{MSRJDReview,Tauber2014,Doi1976a,Doi1976b,Peliti1985}. This mapping yields a set of coupled stochastic partial differential equations (SPDEs) for the physical fields, which we interpret in the Itô sense.

A crucial physical distinction is that cells ($C$, $N$, $T$) typically occur in low copy numbers, where demographic fluctuations are significant, whereas drug molecules ($B$, $D$) are present in large numbers. We therefore treat the drug sector as effectively deterministic, while retaining multiplicative noise in the cellular sector.

\subsection{Chemical sector: deterministic delayed pharmacokinetics}

The PK dynamics follow from the linear terms in the precursor ($\tilde{B}$) and active ($\tilde{D}$) response fields. The precursor field $B(\mathbf{x}, t)$ represents drug in the bloodstream or gastrointestinal tract, whereas $D(\mathbf{x}, t)$ represents active drug in the tumor microenvironment. The corresponding equations of motion are
\begin{align}
   \partial_t B(\mathbf{x}, t) &= \kappa - \alpha_a B(\mathbf{x}, t) ,
   \label{eq:B_dynamics} \\
   \partial_t D(\mathbf{x}, t) &= D_D \nabla^2 D(\mathbf{x}, t)
   + \alpha_a B(\mathbf{x}, t)
   - C_0 D(\mathbf{x}, t) ,
   \label{eq:D_dynamics}
\end{align}
where $\kappa$ is the influx into the precursor compartment, $\alpha_a$ the absorption rate (with characteristic delay $\tau = 1/\alpha_a$), and $C_0$ the decay rate of the active drug. Because fluctuations in $B$ and $D$ scale as $V^{-1/2}$ in the large-volume limit, we neglect their noise contributions and treat Eqs.~\eqref{eq:B_dynamics}–\eqref{eq:D_dynamics} as deterministic  \cite{PKBasics,BeyondDeterministicPK}.

\subsection{Cellular sector: multiplicative Langevin equations}

The tumor and immune fields are driven by the deterministic drift terms generated by the Hamiltonian density $\mathcal{H}_{\mathrm{cell}}$ and by demographic noise \cite{Zygadlo2009,ColouredNoiseRD2020} originating from the quadratic response vertices ($\tilde{C}^2$, $\tilde{N}^2$, $\tilde{T}^2$). The Langevin equation for the tumor density $C(\mathbf{x}, t)$ in the presence of delayed chemotherapy reads
\begin{equation}
    \partial_t C(\mathbf{x}, t)
    = D_C \nabla^2 C
    + a_1 C (1 - a_2 C)
    - a_3 C N
    - \beta_1 C T
    - b_\tau C D
    + \eta_C(\mathbf{x}, t) ,
    \label{eq:C_Langevin}
\end{equation}
while the innate (natural killer) and adaptive (cytotoxic T) immune fields obey
\begin{align}
    \partial_t N(\mathbf{x}, t)
    &= D_N \nabla^2 N
    + a_4 N (1 - a_5 N)
    - a_6 C N
    - \rho N
    - b_\aleph N D
    + \eta_N(\mathbf{x}, t) ,
    \label{eq:N_Langevin} \\
    \partial_t T(\mathbf{x}, t)
    &= D_T \nabla^2 T
    + r C N
    - w T
    - \beta_2 C T
    - b_L T D
    + \eta_T(\mathbf{x}, t) .
    \label{eq:T_Langevin}
\end{align}
Collecting the cellular fields into $\boldsymbol{\Phi} = (C, N, T)^{\mathrm{T}}$, Eqs.~\eqref{eq:C_Langevin}–\eqref{eq:T_Langevin} can be written in compact form as
\begin{equation}
    \partial_t \boldsymbol{\Phi}(\mathbf{x}, t)
    = \mathbf{A}(\boldsymbol{\Phi}, D)
    + \boldsymbol{\eta}(\mathbf{x}, t) ,
    \label{eq:SPDE_vector}
\end{equation}
where $\mathbf{A}(\boldsymbol{\Phi}, D)$ is the deterministic reaction–diffusion drift defined by $\mathcal{H}_{\mathrm{cell}}$ and the cytotoxic couplings to $D$.

\subsection{Noise covariance and multiplicative structure}
\label{sec:noise_covariance}

The stochastic terms $\eta_i(\mathbf{x}, t)$ ($i \in \{C,N,T\}$) represent Gaussian white noises with zero mean,
\begin{equation}
    \langle \eta_i(\mathbf{x}, t) \rangle = 0 ,
\end{equation}
and covariances determined by the underlying Doi–Peliti action \cite{ItakuraOhkuboSasa2009,Lecomte2019}. At the Gaussian (tree) level, the MSRJD mapping of a second-quantized reaction scheme reproduces the Chemical Langevin correspondence: for every elementary reaction channel $r$ with propensity (rate) $a_r(\boldsymbol{\Phi})$ and stoichiometric change $\boldsymbol{\nu}_r$, the noise covariance is
\begin{equation}
    \mathcal{B}_{ij}(\boldsymbol{\Phi})
    = \sum_r a_r(\boldsymbol{\Phi})\, \nu_{r,i}\, \nu_{r,j} .
    \label{eq:CLE_correspondence}
\end{equation}
Inspecting $\mathcal{H}_{\mathrm{cell}}$ in Eq.~\eqref{eq:H_cell} and $\mathcal{L}_{\mathrm{kill}}$ in Eq.~\eqref{eq:L_kill} shows that every vertex carries a single shifted response field, $(\tilde{C}-1)$, $(\tilde{N}-1)$, or $(\tilde{T}-1)$. Physically, this means each elementary event in the model (proliferation, crowding death, drug-induced death, immune-mediated killing, immune depletion, CTL recruitment, or natural decay) changes one cellular species by $\pm 1$, so that $\nu_{r,i}\nu_{r,j}=0$ for $i\neq j$ in every channel. Consequently, from Eq.~\eqref{eq:CLE_correspondence} the off-diagonal covariance vanishes \emph{exactly}, not merely as a subleading approximation:
\begin{equation}
    \mathcal{B}_{CN} = \mathcal{B}_{CT} = \mathcal{B}_{NT} = 0 .
    \label{eq:B_offdiag_zero}
\end{equation}
This is a direct structural consequence of modelling each tumor–immune encounter (e.g.\ NK-mediated killing, tumor-induced NK depletion, CTL recruitment) as its own independent stochastic channel; a model in which a single event simultaneously changed two populations (e.g.\ a joint predator--prey-type event) would instead generate a non-zero cross term $-a_3 CN$ in $\mathcal{B}_{CN}$.

Summing Eq.~\eqref{eq:CLE_correspondence} over \emph{all} channels gives the complete diagonal entries,
\begin{align}
    \langle \eta_C(\mathbf{x}, t) \eta_C(\mathbf{x}', t') \rangle
    &= \left[ a_1 C + a_1 a_2 C^2 + a_3 C N + \beta_1 C T + b_\tau C D \right]
       \delta(\mathbf{x}-\mathbf{x}') \delta(t-t') ,
    \label{eq:CC_cov}\\
    \langle \eta_N(\mathbf{x}, t) \eta_N(\mathbf{x}', t') \rangle
    &= \left[ a_4 N + a_4 a_5 N^2 + a_6 C N + \rho N + b_\aleph N D \right]
       \delta(\mathbf{x}-\mathbf{x}') \delta(t-t') ,
    \label{eq:NN_cov}\\
    \langle \eta_T(\mathbf{x}, t) \eta_T(\mathbf{x}', t') \rangle
    &= \left[ r C N + \beta_2 C T + w T + b_L T D \right]
       \delta(\mathbf{x}-\mathbf{x}') \delta(t-t') .
    \label{eq:TT_cov}
\end{align}
Each bracket collects the shot-noise contribution of \emph{every} reaction channel acting on that species — proliferation and crowding death, immune-mediated killing, mutual depletion, recruitment, natural decay, and drug-induced cytotoxicity alike — consistently with the same rates that already appear in the deterministic drift of Eqs.~\eqref{eq:C_Langevin}--\eqref{eq:T_Langevin}. Retaining only the logistic (proliferation/crowding) part, as in a leading-order demographic-noise approximation, would inconsistently drop terms of the same order as those kept whenever $D$, $N$, or $T$ are not parametrically small.

More generally, we write
\begin{equation}
    \langle \eta_i(\mathbf{x}, t) \eta_j(\mathbf{x}', t') \rangle
    = \mathcal{B}_{ij}(\boldsymbol{\Phi})\,
      \delta(\mathbf{x}-\mathbf{x}') \delta(t-t') ,
    \label{eq:B_def}
\end{equation}
with a noise covariance matrix $\mathcal{B}(\boldsymbol{\Phi})$ that, by Eq.~\eqref{eq:B_offdiag_zero}, is \emph{exactly} diagonal:
\begin{eqnarray}
  && \mathcal{B}(\boldsymbol{\Phi})
    = \nonumber  \\
  && \begin{pmatrix}
        a_1 C + a_1 a_2 C^2 + a_3 C N + \beta_1 C T + b_\tau C D & 0 & 0 \\
        0 & a_4 N + a_4 a_5 N^2 + a_6 C N + \rho N + b_\aleph N D & 0 \\
        0 & 0 & r C N + \beta_2 C T + w T + b_L T D
    \end{pmatrix} .
    \label{eq:B_matrix}
\end{eqnarray}

For analytical and numerical purposes it is convenient to express the noise in terms of independent standard Gaussian fields $\xi_i(\mathbf{x}, t)$, satisfying
\begin{equation}
    \langle \xi_i(\mathbf{x}, t) \xi_j(\mathbf{x}', t') \rangle
    = \delta_{ij} \delta(\mathbf{x}-\mathbf{x}') \delta(t-t') .
\end{equation}
We introduce a noise amplitude matrix $\mathcal{M}(\boldsymbol{\Phi})$ such that
\begin{equation}
    \mathcal{B}(\boldsymbol{\Phi})
    = \mathcal{M}(\boldsymbol{\Phi}) \,
      \mathcal{M}^{\mathrm{T}}(\boldsymbol{\Phi}) ,
    \label{eq:B_MM}
\end{equation}
and choose the diagonal representation
\begin{eqnarray}
  &&  \mathcal{M}(\boldsymbol{\Phi})
    = \nonumber \\
  &&  \begin{pmatrix}
        \sqrt{a_1 C + a_1 a_2 C^2 + a_3 C N + \beta_1 C T + b_\tau C D} & 0 & 0 \\
        0 & \sqrt{a_4 N + a_4 a_5 N^2 + a_6 C N + \rho N + b_\aleph N D} & 0 \\
        0 & 0 & \sqrt{r C N + \beta_2 C T + w T + b_L T D}
    \end{pmatrix} .
    \label{eq:M_matrix}
\end{eqnarray}
Writing $\boldsymbol{\eta} = \mathcal{M}(\boldsymbol{\Phi}) \boldsymbol{\xi}$ and substituting into Eq.~\eqref{eq:SPDE_vector}, we obtain the explicit multiplicative Langevin form
\begin{align}
    \partial_t C(\mathbf{x}, t)
    &= D_C \nabla^2 C
    + a_1 C - a_1 a_2 C^2
    - a_3 C N
    - \beta_1 C T
    - b_\tau C D
    + \sqrt{a_1 C + a_1 a_2 C^2 + a_3 C N + \beta_1 C T + b_\tau C D}\, \xi_C(\mathbf{x}, t) ,
    \label{eq:C_mult} \\
    \partial_t N(\mathbf{x}, t)
    &= D_N \nabla^2 N
    + a_4 N - a_4 a_5 N^2
    - a_6 C N
    - \rho N
    - b_\aleph N D
    + \sqrt{a_4 N + a_4 a_5 N^2 + a_6 C N + \rho N + b_\aleph N D}\, \xi_N(\mathbf{x}, t) ,
    \label{eq:N_mult} \\
    \partial_t T(\mathbf{x}, t)
    &= D_T \nabla^2 T
    + r C N
    - w T
    - \beta_2 C T
    - b_L T D
    + \sqrt{r C N + \beta_2 C T + w T + b_L T D}\, \xi_T(\mathbf{x}, t) .
    \label{eq:T_mult}
\end{align}
These SPDEs provide the starting point for the subsequent Fokker–Planck analysis of micro-cluster survival. Note that, because the noise amplitude in each channel now follows directly and consistently from the same rates entering the drift, the demographic-noise strength correctly tracks the cytotoxic drug terms $b_\tau CD$, $b_\aleph ND$, $b_L TD$: during peak drug exposure the shot noise from drug-induced killing is not negligible, and it is retained here rather than only appearing in the deterministic sink.

\section{Fokker - Planck dynamics and survival probability}
\label{sec:fokker_planck}

To obtain an analytical expression for the relapse probability, we focus on a clinically motivated limit: a residual tumor cell cluster residing in a spatial ``sanctuary site'' near the peak of a chemotherapy cycle. In this regime we make two simplifying assumptions:
\begin{itemize}
    \item The active drug $D(t)$ has driven the local innate and adaptive immune populations to extinction, so that $N = 0$ and $T = 0$.
    \item The surviving tumor density is very small, $C \ll 1/a_2$, so that logistic crowding corrections proportional to $C^2$ are negligible compared to linear demographic terms.
\end{itemize}
Applying these assumptions to the complete covariance of Eq.~\eqref{eq:CC_cov} rather than to the truncated form used previously, the terms $a_3 CN$ and $\beta_1 CT$ vanish identically because $N=T=0$, while $a_1 a_2 C^2$ is negligible relative to $a_1 C$ because $C \ll 1/a_2$. The drug-induced shot-noise term $b_\tau C D$, however, is \emph{not} parametrically small in this regime: $D(t)$ is, by construction, at or near its peak, and $b_\tau C D$ is of the same order in $C$ as the surviving linear term $a_1 C$. It must therefore be retained, giving an effective noise strength
\begin{equation}
    \sigma(t) = a_1 + b_\tau D(t)
    \label{eq:sigma_t}
\end{equation}
in place of the constant $a_1$ used in a leading-order demographic-noise treatment. Under these conditions the tumor field obeys an effectively one-dimensional stochastic differential equation of Feller type \cite{Feller1951,CIR1985},
\begin{equation}
    \partial_t C(t) = \lambda(t)\, C(t)
    + \sqrt{\sigma(t)\, C(t)}\, \xi_C(t) ,
    \label{eq:Feller_SDE}
\end{equation}
with an effective net growth rate
\begin{equation}
    \lambda(t) = a_1 - b_\tau D(t) ,
    \label{eq:lambda_t}
\end{equation}
and Gaussian white noise $\xi_C(t)$ satisfying $\langle \xi_C(t) \rangle = 0$ and $\langle \xi_C(t) \xi_C(t') \rangle = \delta(t-t')$. We work in the Itô convention, consistent with the Doi–Peliti/MSRJD derivation. Note that $\lambda(t)$, which is fixed by the deterministic drift, is unaffected by this correction; only the noise strength $\sigma(t)$ multiplying $C(t)$ changes.

The corresponding Fokker–Planck equation for the probability density $P(C,t \mid C_{\mathrm{ini}},0)$ is
\begin{equation}
    \partial_t P
    = - \partial_C \bigl[\lambda(t)\, C P\bigr]
      + \frac{1}{2} \partial_C^2 \bigl[\sigma(t)\, C P\bigr] ,
    \label{eq:FP_eq}
\end{equation}
with $C \ge 0$ and an absorbing boundary at $C=0$. The survival probability of the micro-cluster is then given by the probability mass in the positive half-line,
\begin{equation}
    P_{\mathrm{surv}}(t)
    = \int_0^\infty P(C,t \mid C_{\mathrm{ini}},0)\, dC .
\end{equation}
Here $C_{\mathrm{ini}}$ denotes the initial size of the surviving micro-cluster, and should not be confused with the active-drug decay rate $C_0$ of Eq.~\eqref{eq:D_dynamics} (Table~\ref{tab:sim_params}); we use the distinct symbol $C_{\mathrm{ini}}$ throughout this section to avoid the notational collision present in an earlier draft.

Solving Eq.~\eqref{eq:FP_eq} by standard methods  \cite{Feller1951,CIR1985,Otunuga2025} yields an exact expression for the survival probability of a cluster starting from $C_{\mathrm{ini}} > 0$:
\begin{equation}
    P_{\mathrm{surv}}(t)
    = 1 - \exp\!\left[
        - \frac{2 C_{\mathrm{ini}}}{\Lambda(t)}
      \right] ,
    \label{eq:P_surv}
\end{equation}
where
\begin{equation}
    \Lambda(t)
    = \int_0^t
      \sigma(s)\,
      \exp\!\left[
        - \int_s^t \lambda(u)\, du
      \right] ds
    \label{eq:Lambda_def}
\end{equation}
is a noise-integrated survival functional that encodes the full time dependence of both the drift $\lambda(t)$ and the noise strength $\sigma(t)$. Equation~\eqref{eq:Lambda_def} reduces to the leading-order demographic-noise result, $\Lambda(t) = a_1\int_0^t \exp[-\int_s^t \lambda(u)\,du]\,ds$, only in the limit $b_\tau D(t) \ll a_1$, which is not satisfied near the peak of a chemotherapy cycle.

The behaviour of $\Lambda(t)$ is controlled by the temporal profile of the drug concentration $D(t)$. $D(t)$ decays on a time scale set by $C_0$, so that  $\lambda(t) = a_1 - b_\tau D(t)$ becomes positive once the chemotherapeutic effect has subsided. When $\lambda(t)$ is strictly negative over an initial time window (during peak drug exposure) but becomes positive at later times, 
 $\Lambda(t)$ approaches a finite, strictly positive constant as $t \to \infty$, and Eq.~\eqref{eq:P_surv} shows that $P_{\mathrm{surv}}(t)$ saturates at a strictly positive constant: 
\begin{equation}
    \lim_{t\to\infty} P_{\mathrm{surv}}(t)
    = 1 - \exp\!\left[
        - \frac{2 C_{\mathrm{ini}}}{\Lambda(\infty)}
      \right] > 0 .
\end{equation}
Thus, for any finite drug influx $\kappa$ and cytotoxic strength $b_\tau$, demographic noise encoded in $\sigma(t)$ guarantees a non-zero probability that a residual tumor cell cluster survives the treatment window. Because the immune fields have been driven to an absorbing state ($N = T = 0$), these surviving cells then experience a positive net growth rate and can undergo rapid regrowth once $D(t)$ has decayed.

\begin{figure}[t]
  \centering
  \includegraphics[width=\linewidth]{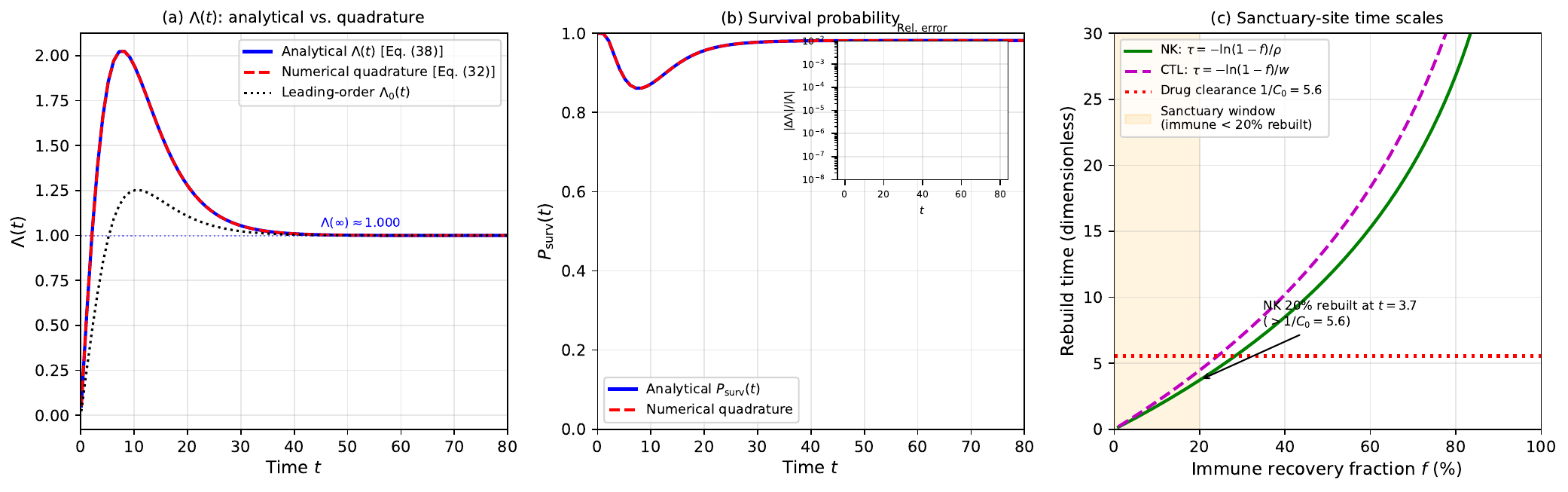}
  \caption{Analytical verification and sanctuary-site justification.
  \textbf{(a)} Closed-form $\Lambda(t)$ (Eq.~\eqref{eq:Lambda_corrected}, blue solid)
  compared with direct numerical quadrature of Eq.~\eqref{eq:Lambda_def} (red dashed)
  and the leading-order approximation $\Lambda_0(t)$ (black dotted), for $D_0=5$,
  $C_0=0.18$, $a_1=0.22$, $b_\tau=0.060$. The asymptotic plateau $\Lambda(\infty)\approx1.000$
  is marked by a horizontal dotted line.
  \textbf{(b)} Corresponding $P_\mathrm{surv}(t)$ (Eq.~\eqref{eq:P_surv}, $C_\mathrm{ini}=2$)
  from both methods. Inset: relative error $|\Delta\Lambda|/|\Lambda|$; the maximum deviation
  is $<3\times10^{-15}$, confirming the exactness of Eq.~\eqref{eq:Lambda_corrected}.
  \textbf{(c)} Sanctuary-site time-scale justification: immune recovery times
  $-\ln(1-f)/\rho$ (NK, green) and $-\ln(1-f)/w$ (CTL, magenta) as functions of the
  recovery fraction $f$, compared to the drug clearance time $1/C_0=5.6$ (red dotted,
  parameters from Table~\ref{tab:sim_params}). The orange shaded region ($f<20\%$) marks
  the sanctuary window in which both immune populations remain substantially depleted,
  confirming that the Feller reduction $N=T=0$ is quantitatively justified for the
  relevant post-bolus interval.}
  \label{fig:lambda_verification}
\end{figure}

To illustrate the behaviour of $\Lambda(t)$, consider a simple bolus protocol in which, after an initial transient, the active drug concentration decays exponentially,
\begin{equation}
    D(t) = D_0 \, e^{-C_0 t} ,
    \label{eq:D_exponential}
\end{equation}
with $D_0 > 0$ and decay rate $C_0 > 0$. In this case one has:
\begin{equation}
    \lambda(t) = a_1 - b_\tau D_0 e^{-C_0 t} , \qquad
    \sigma(t) = a_1 + b_\tau D_0 e^{-C_0 t} .
\end{equation}
$\lambda(t)$ is negative at early times if $b_\tau D_0 > a_1$, but becomes positive once the drug has sufficiently decayed. The inner integral in Eq.~\eqref{eq:Lambda_def} is unchanged from the leading-order treatment, since it depends only on $\lambda(t)$,
\begin{equation}
    \int_s^t \lambda(u)\, du
    = a_1 (t-s)
      - \frac{b_\tau D_0}{C_0}
        \bigl( e^{-C_0 s} - e^{-C_0 t} \bigr) .
    \label{eq:inner_integral}
\end{equation}
It is convenient to define the leading-order functional
\begin{equation}
    \Lambda_0(t)
    = a_1 \int_0^t
      \exp\!\left\{
        - a_1 (t-s)
        + \frac{b_\tau D_0}{C_0}
          \bigl( e^{-C_0 s} - e^{-C_0 t} \bigr)
      \right\} ds ,
    \label{eq:Lambda_exponential}
\end{equation}
which coincides with the survival functional obtained when the noise strength is held fixed at $\sigma = a_1$. Writing $\varphi(s) \equiv (b_\tau D_0/C_0)\,e^{-C_0 s}$, one has $b_\tau D_0 e^{-C_0 s} = -\varphi'(s)$, so the additional contribution from the drug-dependent part of $\sigma(s)$ in Eq.~\eqref{eq:Lambda_def} can be integrated by parts in closed form. This yields the exact result
\begin{equation}
    \Lambda(t)
    = 2\, \Lambda_0(t) - 1
      + \exp\!\left[
          -a_1 t + \frac{b_\tau D_0}{C_0}\bigl(1-e^{-C_0 t}\bigr)
        \right] ,
    \label{eq:Lambda_corrected}
\end{equation}
which we have verified against direct numerical quadrature of Eq.~\eqref{eq:Lambda_def}. Fig.~\ref{fig:lambda_verification}(a,b) shows this comparison explicitly: the maximum relative deviation between the closed-form expression and the quadrature result is $|\Delta\Lambda|/|\Lambda| < 3\times10^{-15}$ across all $t$ values used in the simulations. Because $\Lambda(t) > \Lambda_0(t)$ at all finite $t$ (the additional drug-induced shot noise strictly increases the noise-integrated functional), the corrected survival probability $P_{\mathrm{surv}}(t)$ in Eq.~\eqref{eq:P_surv} is strictly \emph{smaller} than the leading-order estimate at finite times, since $P_{\mathrm{surv}}$ is a decreasing function of $\Lambda$: stronger demographic noise accelerates absorption at $C=0$ rather than delaying it. The two estimates coincide only in the asymptotic limit $t\to\infty$, where the drug-dependent excess noise has decayed away and $\Lambda(\infty) = 2\Lambda_0(\infty)-1$.

For large times $t \gg C_0^{-1}$ the term $e^{-C_0 t}$ is negligible, and one has:
\begin{equation}
    \exp\!\left[
        - a_1 (t-s)
        + \frac{b_\tau D_0}{C_0} e^{-C_0 s}
      \right]
    \xrightarrow[t\to\infty]{}
    e^{-a_1 (t-s)} ,
\end{equation}
so that the integral in Eq.~\eqref{eq:Lambda_exponential} converges to a constant of order $1/a_1$ as $t-s \to \infty$ when $t$ is going to $\infty$. Consequently $\Lambda_0(t)$, and via Eq.~\eqref{eq:Lambda_corrected} also the corrected $\Lambda(t)$, approach finite positive constants as $t \to \infty$, and the survival probability \eqref{eq:P_surv} saturates at a positive value, in agreement with the qualitative discussion above. The quantitative value of that plateau, and the transient approach to it, are shifted downward relative to the leading-order estimate by the mechanism identified above; the numerical results of Figs.~\ref{fig:feller_three_schedules}--\ref{fig:feller_kappa_corrected} (Sec.~\ref{sec:topological}) are obtained with the corrected noise strength $\sigma(t)$ from Eq.~\eqref{eq:sigma_t} throughout, and the threshold condition of Eq.~\eqref{eq:kappa_c} remains unaffected, since it depends only on the sign of $\lambda^*$.

\begin{figure}
    \centering
    \includegraphics[width=\linewidth]{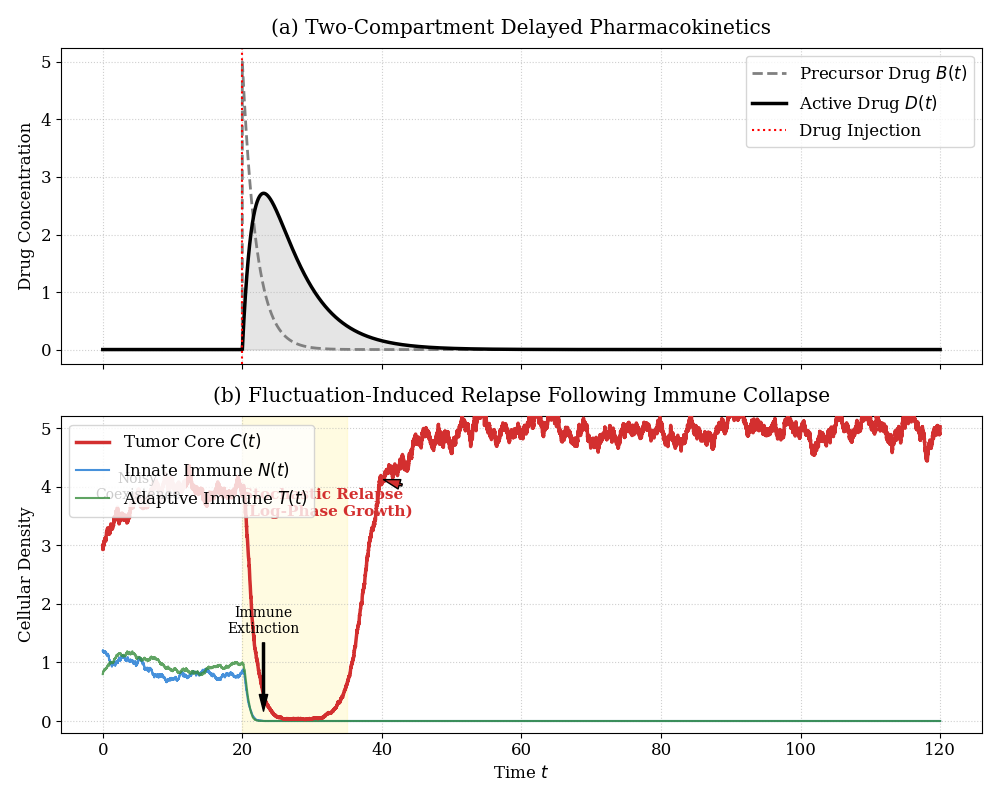}
    \caption{Simulated dynamics of the tumor microenvironment under delayed chemotherapy. 
    \textbf{(a)} Two-compartment pharmacokinetics: a bolus injection of dose $D_{\mathrm{dose}} = 5.0$ at $t = 20$ into the precursor field $B(t)$ is absorbed into the active tissue field $D(t)$ at rate $\alpha_a = 0.5$ and decays with rate $C_0 = 0.2$. 
    \textbf{(b)} Corresponding population dynamics for tumor and immune fields (parameters are fixed to $a_1=1$, $a_2=0.2$, $a_3=0.15$, $a_4=0.8$, $a_5=0.5$, $a_6=0.1$, $\rho=0.1$, $r=0.15$, $w=0.1$, $\beta_1=0.1$, $\beta_2=0.1$, reference carrying capacity $\Omega=C^\ast+N^\ast+T^\ast=200.0$ as defined in Eq.~\eqref{eq:Omega_def}). 
    During the active drug window (shaded region), differential cytotoxicity ($b_\tau=0.65$, $b_\aleph=1.3$, $b_L=1.3$) strongly depletes the innate ($N$) and adaptive ($T$) immune populations, while demographic fluctuations allow a small tumor micro-cluster to persist. 
    As $D(t)$ decays, the surviving cluster regrows in an immune-depleted background, illustrating the stochastic relapse mechanism predicted by the field-theoretic model.}
    \label{fig:stochastic_relapse}
\end{figure}

\section{Topological interventions: minimizing the survival functional}
\label{sec:topological}

The analytical form of $P_{\mathrm{surv}}(t)$ in Eqs.~\eqref{eq:P_surv}–\eqref{eq:Lambda_def} shows that relapse control requires modifying either the effective drift $\lambda(t)$ or the absorbing structure of the immune fields. In this section we propose two intervention strategies suggested by the stochastic theory.

\subsection{Continuous or metronomic dosing: flattening the PK profile}

After an initial period with $\lambda(t) < 0$, the drug concentration decreases, $\lambda(t)$ becomes positive, and $\Lambda(t)$ diverges,inducing  a  survival probability. A natural way to avoid this regime is to replace a single bolus by a sustained influx $\kappa > 0$ into the precursor compartment \cite{Browder2000Metronomic,MetronomicReview2014,MetronomicHematology2024}.

For a constant infusion, the PK equations
\begin{equation}
    \partial_t B = \kappa - \alpha_a B, \qquad
    \partial_t D = D_D \nabla^2 D + \alpha_a B - C_0 D ,
\end{equation}
relax to a homogeneous steady state with
\begin{equation}
    B^* = \frac{\kappa}{\alpha_a}, \qquad
    D^* = \frac{\alpha_a B^*}{C_0} = \frac{\kappa}{C_0} .
\end{equation}
In this regime $\lambda(t)$ becomes time-independent,
\begin{equation}
    \lambda^* = a_1 - b_\tau D^*
               = a_1 - b_\tau \frac{\kappa}{C_0} .
\end{equation}
If the infusion rate $\kappa$ is chosen such that
\begin{equation}
    b_\tau \frac{\kappa}{C_0} > a_1 ,
    \label{eq:neg_drift_condition}
\end{equation}
then $\lambda^* < 0$ for all $t$. Into Eq.~\eqref{eq:Lambda_def} this  yields a finite $\Lambda(\infty)$, and the survival probability \eqref{eq:P_surv} decays to zero. Thus, in the idealized limit of a perfectly maintained infusion that satisfies Eq.~\eqref{eq:neg_drift_condition}, the stochastic window of vulnerability is eliminated.

\begin{figure}[t]
  \centering
  \includegraphics[width=\linewidth]{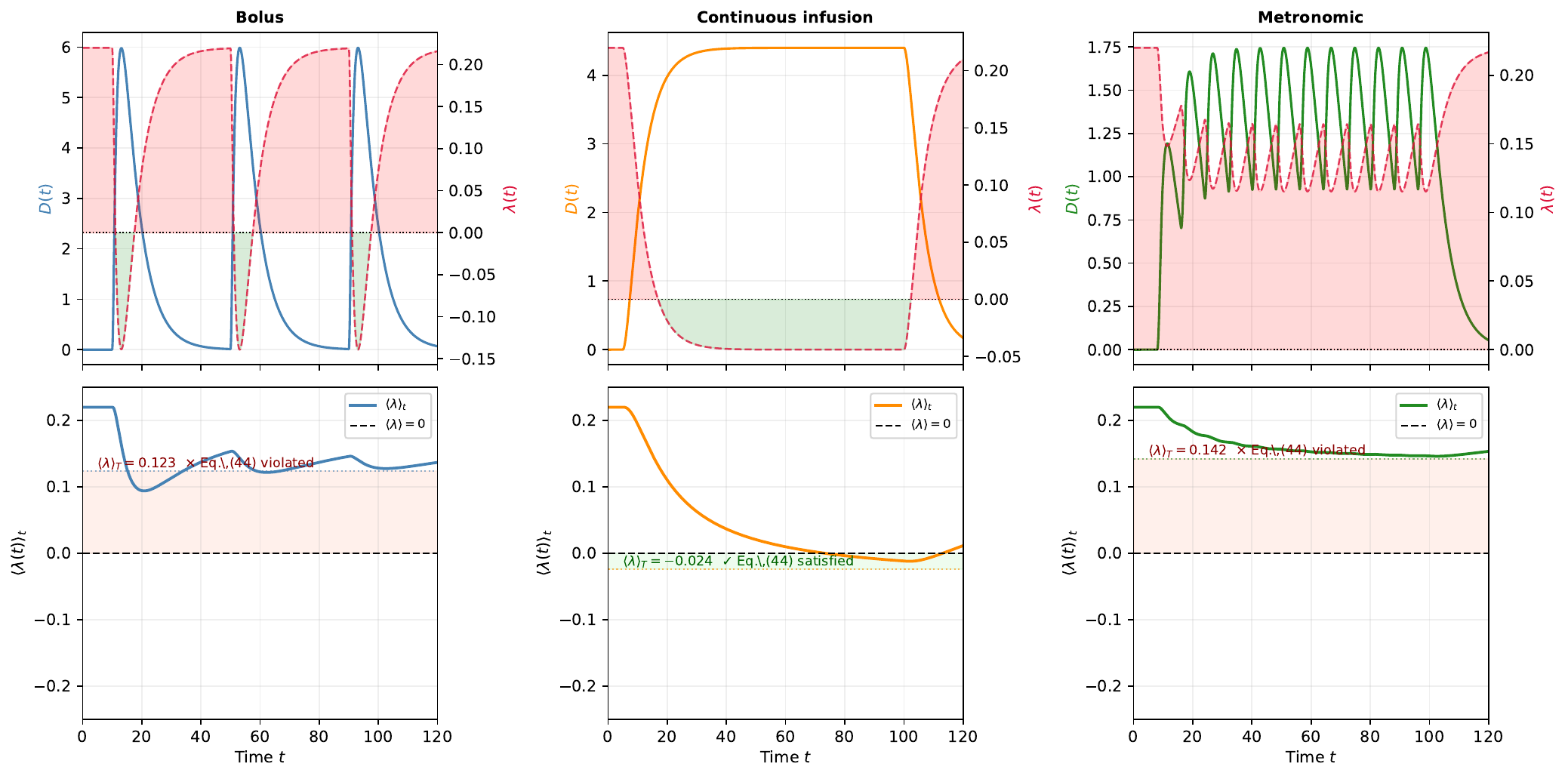}
  \caption{Numerical evaluation of the time-averaged drift condition
  for each dosing protocol.
  \textbf{Top row:} Active drug concentration $D(t)$ (left axis, colored solid) and
  effective drift $\lambda(t)=a_1-b_\tau D(t)$ (right axis, crimson dashed) for the
  bolus (left), continuous infusion at $\kappa=1.2\,\kappa_c$ (centre), and metronomic
  (right) protocols. Green shading: $\lambda(t)<0$ (tumor suppressed); red shading:
  $\lambda(t)>0$ (net tumor growth).
  \textbf{Bottom row:} Running time-average $\langle\lambda\rangle_t =
  t^{-1}\int_0^t\lambda(s)\,ds$. Light green (resp.\ salmon) background indicates
  that the final value satisfies (resp.\ violates) Eq.~(44).
  The continuous infusion arm achieves $\langle\lambda\rangle_T=-0.024<0$, confirming
  that the stochastic window of vulnerability is eliminated for $\kappa>\kappa_c$,
  while both the bolus and metronomic protocols leave a positive time-averaged drift
  and correspondingly high relapse probabilities.}
  \label{fig:metronomic_condition}
\end{figure}

In practice, continuous infusion is approximated by metronomic or low-dose pulsed schedules. From the perspective of Eq.~\eqref{eq:Lambda_def}, such protocols should be designed so that the time-averaged effective drift remains negative,
\begin{equation}
    \frac{1}{T} \int_0^T \lambda(t)\, dt < 0
\end{equation}
over intervals comparable to the characteristic extinction time of residual tumor cell clusters, thereby keeping $\Lambda(t)$ bounded and suppressing $P_{\mathrm{surv}}$ \cite{Browder2000Metronomic,MetronomicReview2014}. Fig.~\ref{fig:metronomic_condition} evaluates this condition numerically for the three protocols of Sec.~\ref{sec:immunotherapy_numerics}: the continuous infusion arm ($\kappa=1.2\,\kappa_c$) achieves $\langle\lambda\rangle_T = -0.024 < 0$ (condition satisfied), whereas both the bolus ($\langle\lambda\rangle_T = +0.12$) and the metronomic arm ($\langle\lambda\rangle_T = +0.14$) violate it, consistent with their elevated relapse probabilities ($P_\mathrm{relapse}=0.028$ and $0.548$, respectively).

\subsection{Adjuvant immunotherapy: breaking the immune absorbing state}

When high-dose cytotoxic regimens are required, the stochastic dynamics indicate a different vulnerability: the immune fields $N$ and $T$ can be driven into absorbing states, leaving surviving tumor clusters in an  immune-depleted background \cite{ImmuneReconstitution2023}. In order to avoid this negative effect of the medical treatment, we introduce source terms that repopulate the immune sector after the drug has cleared.

At the level of the Langevin equations \eqref{eq:N_mult}–\eqref{eq:T_mult}, this corresponds to
\begin{align}
    \partial_t N &= \dots + \gamma_N(t-t_{\mathrm{delay}}) , \\
    \partial_t T &= \dots + \gamma_T(t-t_{\mathrm{delay}}) ,
    \label{eq:gamma_N_source}
\end{align}
where $\gamma_N$ and $\gamma_T$ represent controlled influxes (e.g.\ adoptive cell transfer or cytokine stimulation) \cite{RebuildImmunity2005,AdoptiveTcellReview}, applied after a delay $t_{\mathrm{delay}}$ chosen such that the active drug concentration $D(t)$ is sufficiently small to avoid the suppression of the new immune cell population. These terms convert the  absorbing boundary at $N=T=0$ into a driven state with a non-zero stationary density, thereby restoring the predation term $-a_3 C N$ in the tumor dynamics during the critical post-treatment period.

From the viewpoint of the survival functional, early immune repopulation modifies the effective drift experienced by a residual cluster from $\lambda(t)$ to
\begin{equation}
    \lambda_{\mathrm{eff}}(t)
    = a_1 - b_\tau D(t) - a_3 N(t) - \beta_1 T(t) ,
    \label{eq:lambda_eff}
\end{equation}
so that, provided the sources $\gamma_N$ and $\gamma_T$ are strong enough to rebuild $N(t)$ and $T(t)$ before $\lambda_{\mathrm{eff}}(t)$ becomes positive, the growth of $\Lambda(t)$ can again be curtailed and $P_{\mathrm{surv}}$ reduced \cite{George2018StochasticImmune}.


\subsection{Connection to dosing scales and immune sources}

For continuous or metronomic chemotherapy \cite{MetronomicBridge2024,MetronomicSystematic2019,ChemoScheduleStochastic2017}, a low,  constant oral or infusion doses is given to the patient in order to maintain drug exposure over long periods, in contrast to intermittent maximum tolerated dose regimens. This kind of protocols corresponds to choosing an infusion rate $\kappa$ that maintains $D(t)$ near a quasi-steady level $D^* \simeq \kappa/C_0$, while respecting toxicity constraints:
\begin{equation}
    b_\tau \frac{\kappa}{C_0} \gtrsim a_1
    \label{eq:clinical_kappa_condition}
\end{equation}
In such a case,  the survival functional $\Lambda(t)$ is to stay bounded.
 Although $a_1$ and $b_\tau$ are not observable, they can in principle be calibrated from longitudinal tumor size or biomarker data under given regimens, allowing Eq.~\eqref{eq:clinical_kappa_condition} to be interpreted as a constraint on feasible metronomic schedules.

For immunotherapy, adoptive T-cell transfer and cytokine-based stimulation protocols deliver controlled numbers of effector cells or immune-stimulating agents over defined time windows \cite{AdoptiveTcellReview,ImmuneReconstitution2021}. In our effective description, these correspond to source terms $\gamma_N(t)$ and $\gamma_T(t)$ that rebuild $N(t)$ and $T(t)$ after chemotherapy. A minimal requirement to  alter the survival probability is that the induced immune densities become comparable to the tumor scale before the $\lambda(t)$ turns positive. In terms of the effective drift, one gets
\begin{equation}
    \lambda_{\mathrm{eff}}(t)
    = a_1 - b_\tau D(t) - a_3 N(t) - \beta_1 T(t) ,
\end{equation}
this suggests designing schedules such that, for times $t$  following drug clearance, one has
\begin{equation}
    a_3 N(t) + \beta_1 T(t) \gtrsim a_1 ,
    \label{eq:clinical_gamma_condition}
\end{equation}
ensuring that $\lambda_{\mathrm{eff}}(t)$ remains non-positive during the critical post-treatment period forbidding   micro-clusters to escape. 
\section{Results and discussion}

The stochastic field-theoretic formulation developed above yields three main qualitative results. First, the two-compartment PK sector coupled to the tumor–immune fields produces a finite absorption delay during which the tumor can continue to fluctuate and grow before the full cytotoxic effect of the drug is expressed. Second, the microscopic Doi–Peliti structure implies multiplicative demographic noise for the cellular populations, this noise generates a non-zero survival probability for residual tumor clusters even under high-dose chemotherapy \cite{Mansour2022SDE,Sardanyes2018}. 

Figure~\ref{fig:stochastic_relapse} illustrates the  dynamics for a representative bolus protocol. The precursor and active drug fields $B(t)$ and $D(t)$ exhibit a sharp rise followed by an exponential decay, consistent with Eqs.~\eqref{eq:B_dynamics}–\eqref{eq:D_dynamics}. The tumor and immune fields fluctuate around a metastable coexistence state prior to treatment, as expected from the stochastic reaction–diffusion dynamics. During the active drug window, the cytotoxic terms proportional to $b_\tau$, $b_\aleph$ and $b_L$  deplete both tumor and immune cells, but the  noise terms allow small tumor cell clusters to survive in spatial regions of reduced drug exposure. As $D(t)$ decays, these residual clusters regrow in an immune-depleted environment.

The reduction to a one-dimensional Feller process in a sanctuary site [Eqs.~\eqref{eq:Feller_SDE}–\eqref{eq:FP_eq}] provides an analytical characterization of this behaviour. The survival probability
\begin{equation}
    P_{\mathrm{surv}}(t)
    = 1 - \exp\!\left[ - 2 C_{\mathrm{ini}} / \Lambda(t) \right]
\end{equation}
depends on the time-integrated noise-to-drift functional $\Lambda(t)$ of Eq.~\eqref{eq:Lambda_def}, which in turn encodes the full pharmacokinetic profile through both $\lambda(t) = a_1 - b_\tau D(t)$ and $\sigma(t)=a_1+b_\tau D(t)$, consistent with delayed-effect PK--PD models of tumor growth inhibition \cite{DelayTGI2014}. For a decaying $D(t)$, $\Lambda(t)$ approaches a finite positive plateau $\Lambda(\infty)$, and $P_{\mathrm{surv}}(t)$ saturates at the  value $1-\exp[-2C_{\mathrm{ini}}/\Lambda(\infty)]$.

These stochastic relapse probabilities can be reduced using the same formalism to describe them. Continuous or metronomic dosing schemes that maintain $D(t)$ close to a steady value $D^* \simeq \kappa / C_0$ can, in principle, enforce a strictly negative drift $\lambda^* = a_1 - b_\tau \kappa/C_0 < 0$, thereby keeping $\Lambda(t)$ finite and driving $P_{\mathrm{surv}}(\infty)$ to zero. When such idealized control is not possible, adjuvant immunotherapy terms $\gamma_N$ and $\gamma_T$ that repopulate the immune sector after chemotherapy  shift the net drift from $\lambda(t)$ to $\lambda_{\mathrm{eff}}(t) = a_1 - b_\tau D(t) - a_3 N(t) - \beta_1 T(t)$, providing an additional mechanism to prevent the drift from becoming positive during the critical post-treatment period. These strategies arise  from the structure of the SPDEs and the Fokker–Planck solution, rather than from ad hoc assumptions.


\subsection{Framework novelty}
\label{sec:positioning}

In the literature,  the Stochastic Pharmacokinetic Escape has been studied following three directions.

\emph{(i) Deterministic PK--PD and tumor-growth models.} Classical ODE-based tumor--drug and tumor--immune models \cite{Yin2019,Debnath2025,Wahbi2024,Jarrett2018,mca30060119,10.1371/journal.pcbi.1009822} predict that a sufficiently high cytotoxic dose drives the tumor population to  zero in finite or infinite time. The extinction in a deterministic ODE is absolute.

\emph{(ii) Stochastic tumor and tumor--immune models.}  They have been extended including  demographic noise
\cite{Mansour2022SDE,54213ede0cfd4949b1c8183fbe25e57a,Al-Mekhlafi31122025,Sardanyes2018,GeorgeLevine2018,George2018StochasticImmune,StochasticTumorImmune2019,zhong2005stochasticresonancegrowthtumor,zhong2005influencecorrelatednoisesgrowth}, and shows that such noise can prevent extinction or generate noise-induced persistence and bistability. The drug is  treated as an instantaneous parameter or a simple time-independent kill rate, without an  absorption-delayed pharmacokinetic sector, and the noise amplitude is not derived   from a microscopic reaction scheme.

\emph{(iii) Stochastic and delayed PK/PD models.} A separate body of work incorporates stochasticity or delay directly into pharmacodynamic models of tumor growth inhibition \cite{Truongetal2025,CSIAMTLS2025,Otunuga2025,Lavietal2012,Butner2021,morgan2024existencestabilityoptimaldrug,DelayTGI2014,Qi2005StochasticPKPD,ChemoScheduleStochastic2017}, 
The pharmacokinetic delay is described but they do not derive the demographic noise from an underlying field-theoretic action.

The present work combines the following elements into a single framework (see Table~\ref{tab:novelty_comparison} for a compact comparison with the closest prior work): 
\begin{itemize}
\item  a  two-compartment pharmacokinetic sector coupled to the tumor--immune fields within the same Doi--Peliti action, so that absorption delay and demographic noise share a common microscopic origin rather than being added separately; 
\item a demographic noise covariance derived directly from the reaction vertices via the Chemical Langevin correspondence of Eq.~\eqref{eq:CLE_correspondence};
\item a  mechanism — immune-field absorption under high-dose cytotoxicity  that couples the size of the stochastic escape window to the pharmacokinetic protocol itself; and 
\item a closed-form survival functional $P_{\mathrm{surv}}(t)$, Eq.~\eqref{eq:P_surv}, that is an explicit functional of the dosing schedule $D(t)$ and, via $\lambda_{\mathrm{eff}}(t)$, of any adjuvant immune-repopulation protocol, giving quantitative dosing and immunotherapy-timing criteria (Eqs.~\eqref{eq:kappa_c} and \eqref{eq:clinical_gamma_condition}).
The numerical experiment of Sec.~\ref{sec:immunotherapy_numerics} illustrates  the importance  that the benefit of adjuvant immunotherapy is time-critical relative to the pharmacokinetic decay.
\end{itemize}
\begin{table}[t]
  \centering
  \caption{Comparison of the present framework with the three nearest classes of prior work.
  $\checkmark$ = present; $\circ$ = partial; -- = absent.}
  \begin{tabular}{lcccc}
    \hline
    Feature & Det.\ PK--PD & Stoch.\ tumor--immune & Stoch.\ PK/PD & \textbf{This work} \\
    \hline
    Two-compartment PK delay & $\checkmark$ & -- & $\circ$ & $\checkmark$ \\
    Demographic noise (cellular) & -- & $\checkmark$ & $\circ$ & $\checkmark$ \\
    Noise from reaction vertices & -- & $\circ$ & -- & $\checkmark$ \\
    Immune absorbing-state mechanism & -- & $\circ$ & -- & $\checkmark$ \\
    Closed-form $P_\mathrm{surv}(t)$ & -- & $\circ$ & $\circ$ & $\checkmark$ \\
    Dosing/timing design criteria & $\circ$ & -- & $\circ$ & $\checkmark$ \\
    \hline
  \end{tabular}
  \label{tab:novelty_comparison}
\end{table}

Works such as Refs.~\cite{Mansour2022SDE,Sardanyes2018,GeorgeLevine2018} show  that demographic noise prevents guaranteed extinction. The contribution of our paper is to show  how pharmacokinetic delay and immune-field absorption jointly reopen and control the size of that stochastic window, and to make the resulting survival probability a  testable function of the dosing protocol.

\subsection{Numerical illustration of stochastic relapse across dosing schedules}

In this section, we performed numerical simulations of the full PK–PD SPDE system and of the reduced one-dimensional Feller SDE under three representative dosing schemes: 
\begin{itemize}
    \item  a high-dose bolus protocol,
    \item a continuous infusion approximating ideal metronomic delivery, and 
    \item a low-dose metronomic schedule.
\end{itemize}
For each protocol, the pharmacokinetic sector was integrated according to Eqs.~\eqref{eq:B_dynamics}–\eqref{eq:D_dynamics} with the  influx $\kappa(t)$, while the cellular sector was evolved using an Euler–Maruyama discretization of the multiplicative Langevin equations \eqref{eq:C_mult}–\eqref{eq:T_mult}.

\begin{figure}[t]
  \centering
  \includegraphics[width=\textwidth]{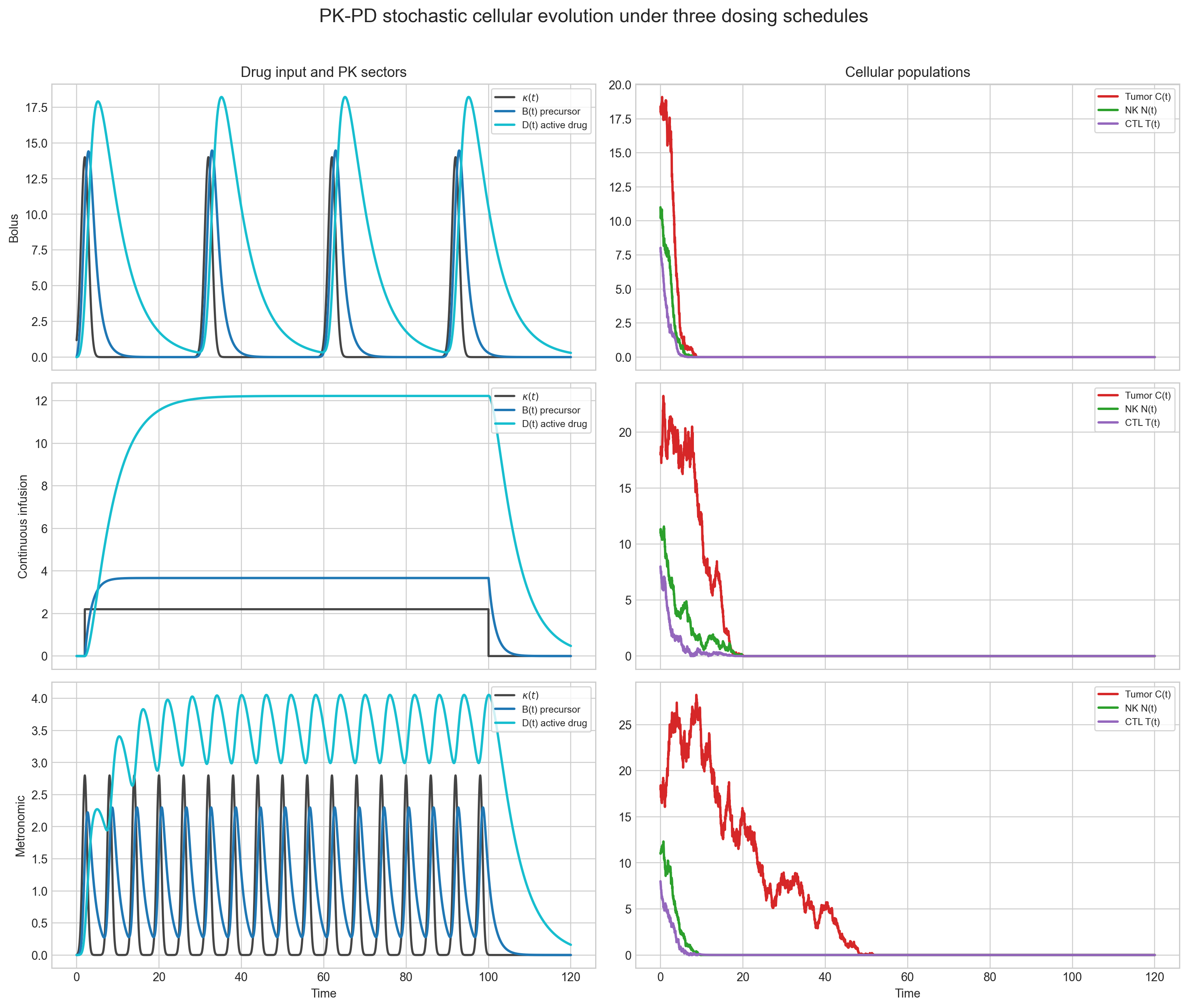}
  \caption{Stochastic PK--PD simulations of the full tumor--immune SPDE system
  with multiplicative demographic noise, using the parameters listed in
  Table~\ref{tab:sim_params}. Left column: pharmacokinetic profiles for
  bolus, continuous infusion, and metronomic dosing. The influx $\kappa(t)$ is
  applied to the precursor compartment $B(t)$ and transferred to the active
  drug $D(t)$ with rate $\alpha_a$, while $D(t)$ decays with rate $C_0$.
  Right column: corresponding sample trajectories of tumor cells $C(t)$ (red),
  NK cells $N(t)$ (green), and CTLs $T(t)$ (purple). All three protocols
  transiently suppress the cellular populations, but differ in the immune depletion and in the residual stochastic fluctuations of the
  tumor density.}
  \label{fig:pkpd_three_schedules}
\end{figure}

Figure~\ref{fig:pkpd_three_schedules} (left panels) shows the resulting PK profiles. In the bolus case, $\kappa(t)$ produces sharp precursor peaks $B(t)$. Under continuous infusion,  $B(t)$ and $D(t)$ rise to quasi-steady plateaus over the treatment window, in agreement with the steady-state values $B^* = \kappa/\alpha_a$ and $D^* = \kappa/C_0$. In the metronomic scheme, a train of low-amplitude pulses yields an oscillatory $D(t)$ that fluctuates around a lower effective mean concentration. The right panels of Fig.~\ref{fig:pkpd_three_schedules}) show that all three protocols can  suppress $C(t)$, $N(t)$, and $T(t)$, but differ in their effects on  immune depletion and on the residual  tumor density.

We then simulated the reduced stochastic equation
\begin{equation}
    \partial_t C(t) = \lambda(t) C(t) + \sqrt{\sigma(t) C(t)}\,\xi(t),
\end{equation}
with $\lambda(t) = a_1 - b_\tau D(t)$ and $\sigma(t) = a_1+b_\tau D(t)$. For each schedule we generated an ensemble of stochastic sample paths starting from a small initial micro-cluster and imposed an absorbing boundary at $C = 0$.  A trajectory is counted as a relapse only if it crosses the relapse threshold $C = 1$ at some intermediate time  and it remains alive at the final observation time $T_{\mathrm{end}}$. 

Figure~\ref{fig:feller_three_schedules} shows the resulting pharmacokinetic profiles and representative stochastic trajectories for the three schedules. In the bolus case, the high transient exposure  suppresses many trajectories and most sample paths fall into the absorbing state or remain subcritical. Under continuous infusion, the  stationary drug concentration keeps the drift negative for  large $\kappa$, reducing both survival and sustained relapse.  In the metronomic case the oscillatory drug profile creates repeated intervals in which the effective drift is  negative or becomes slightly positive, allowing a visible fraction of paths to cross the relapse threshold.

The dependence of the relapse probability on the continuous infusion rate $\kappa$ is shown in Fig.~\ref{fig:feller_kappa_corrected}. As expected,  the terminal survival probability and the corrected relapse probability decrease as $\kappa$ increases and the numerical transition occurs close to:
\begin{equation}
    \kappa_c \simeq \frac{a_1 C_0}{b_\tau},
    \label{eq:kappa_c}
\end{equation}

\subsection{Numerical illustration of adjuvant immunotherapy}
\label{sec:immunotherapy_numerics}

Repopulating the immune compartment after chemotherapy shifts the effective drift experienced by a residual micro-cluster from $\lambda(t)$ to $\lambda_{\mathrm{eff}}(t) = a_1 - b_\tau D(t) - a_3 N(t) - \beta_1 T(t)$, and correspondingly the noise strength from $\sigma(t)=a_1+b_\tau D(t)$ to $\sigma_{\mathrm{eff}}(t) = a_1 + b_\tau D(t) + a_3 N(t) + \beta_1 T(t)$. We test this prediction  by simulating the reduced Feller equation
\begin{equation}
    \partial_t C(t) = \lambda_{\mathrm{eff}}(t)\, C(t) + \sqrt{\sigma_{\mathrm{eff}}(t)\, C(t)}\;\xi(t)
    \label{eq:Feller_immuno}
\end{equation}
for a micro-cluster surviving a representative bolus dose ($D_0=5$, $C_0=0.18$), under two scenarios: 
\begin{enumerate}
    \item  no adjuvant intervention, for which $N(t)=T(t)=0$ for all $t$ as in the baseline sanctuary-site assumption; and
    \item  adjuvant immunotherapy, in which constant source terms $\gamma_N, \gamma_T$ are switched on at $t_{\mathrm{delay}}=6$ — when the deterministic drift $\lambda(t)$ first turns positive — and drive $N(t)$ and $T(t)$ from $0$ towards quasi-steady values $N^\ast \simeq \gamma_N/\rho \approx 15$ and $T^\ast \simeq \gamma_T/w \approx 10$, comparable to the pre-treatment homeostatic levels of Table~\ref{tab:sim_params}.
\end{enumerate}

\begin{figure}[t]
    \centering
    \includegraphics[width=\textwidth]{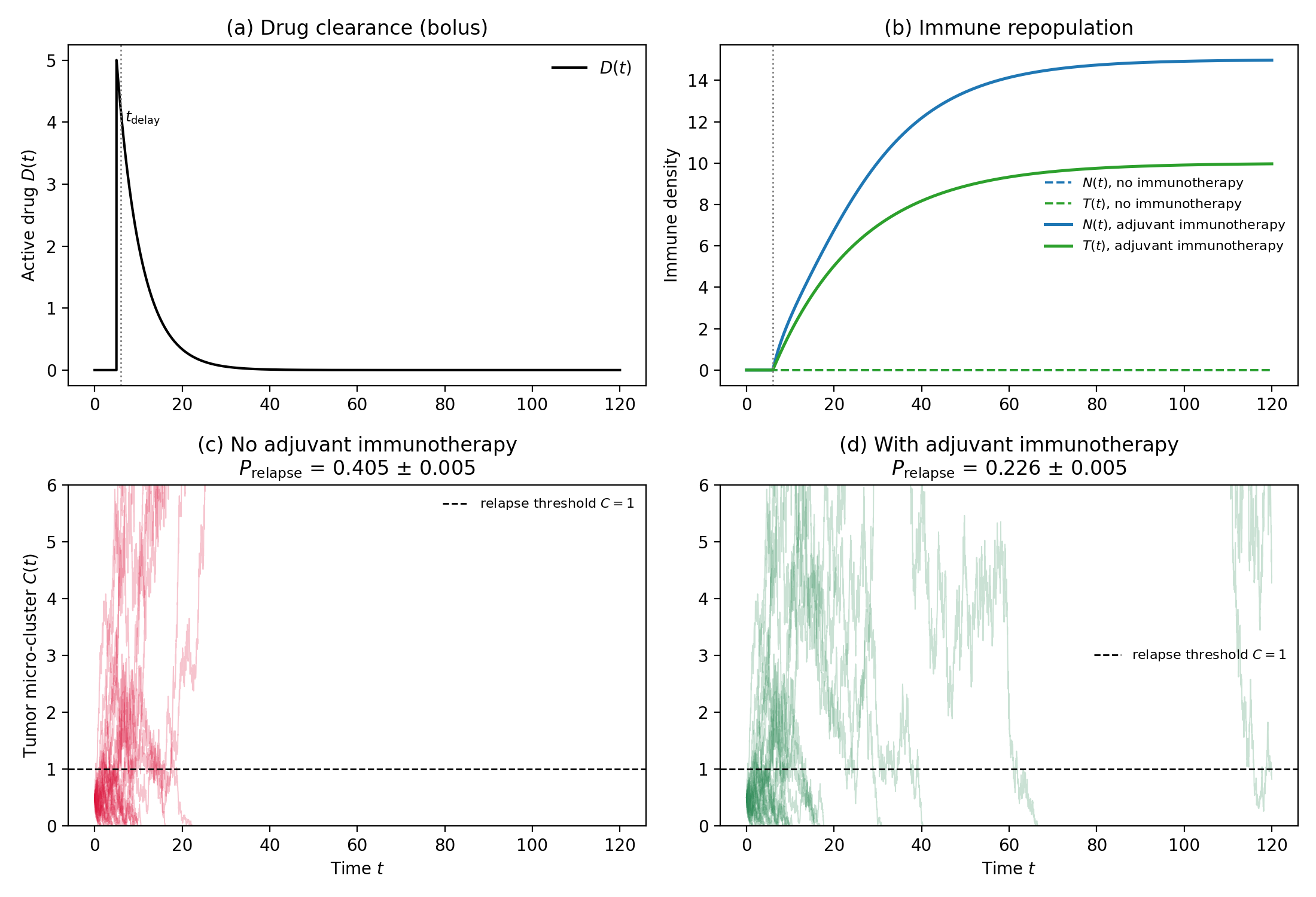}
    \caption{Numerical test of adjuvant immunotherapy in the reduced Feller description, Eq.~\eqref{eq:Feller_immuno}. \textbf{(a)} Active drug concentration $D(t)$ for a representative bolus dose; $t_{\mathrm{delay}}$ marks the onset of adjuvant immune repopulation. \textbf{(b)} Immune trajectories $N(t)$, $T(t)$ with (solid) and without (dashed, identically zero) adjuvant immunotherapy. \textbf{(c,d)} Bundles of $n=30$ sample paths (out of $n=8000$ used to estimate $P_{\mathrm{relapse}}$) of the tumor micro-cluster $C(t)$ without (c) and with (d) adjuvant immunotherapy. Without intervention, $P_{\mathrm{relapse}}=0.405\pm0.005$: surviving clusters escape almost entirely within the early post-bolus window. With adjuvant immunotherapy, $P_{\mathrm{relapse}}=0.226\pm0.005$, a reduction of about $44\%$: the immune source terms cannot prevent the earliest stochastic escapes, which occur before $N(t), T(t)$ have had time to build up, but they substantially increase the extinction rate of clusters that survive into the intermediate-to-late window, consistent with the predation terms $a_3 C N$ and $\beta_1 C T$ acting on $\lambda_{\mathrm{eff}}(t)$.}
    \label{fig:immunotherapy_numerics}
\end{figure}

The results, shown in Fig.~\ref{fig:immunotherapy_numerics}, confirm the qualitative prediction of Sec.~\ref{sec:topological} but we can note that  because immune repopulation is itself described by the natural decay rates $\rho$ and $w$, adjuvant immunotherapy is most effective at suppressing \emph{late} relapse  and comparatively ineffective against clusters that escape stochastically within the first few time units after the drug peak, before $N(t)$ and $T(t)$ have risen appreciably.So, in practice, the benefit of adjuvant immunotherapy for suppressing Stochastic Pharmacokinetic Escape depends on how promptly immune reconstitution begins relative to the pharmacokinetic decay time scale $1/C_0$, a prediction that could in principle be tested against the timing of adoptive T-cell transfer or cytokine support protocols relative to the chemotherapy cycle.

\subsection{Empirical signature of relapse-after-response in public clinical data}
\label{sec:real_data}

We examined the fully anonymized, publicly released longitudinal lesion-volume dataset of  \cite{10.1371/journal.pcbi.1009822}, comprising 1461 patients from five clinical trials of chemotherapy (docetaxel) and immunotherapy (atezolizumab) in non-small-cell lung and bladder cancer, with repeated CT-based target-lesion measurements per patient. 

Using a $20\%$ relative regrowth threshold from the nadir, $60.8\%$ of evaluable patients show this nadir-then-regrowth pattern. It ranges from $65.1\%$ (at a $10\%$ regrowth threshold) to $49.0\%$ (at a strict $50\%$ threshold).
The pattern is also not concentrated in one arm or trial: across the 14 anonymized study/arm codes, the fraction ranges narrowly between $50\%$ and $65\%$ (Table~\ref{tab:kather_summary}), consistent with a broad clinical phenomenon. Among patients showing the pattern, the median time to nadir is $50$ days after the first measurement, with a median of $163$ further days of observed regrowth. 

\begin{table}[t]
  \centering
  \caption{Sensitivity of the nadir-then-regrowth fraction to the regrowth threshold in the reference  \cite{10.1371/journal.pcbi.1009822} dataset ($n=1064$ patients with $\geq 4$ lesion measurements).}
  \begin{tabular}{lc}
    \hline
    Criterion & Fraction showing pattern \\
    \hline
    Volume, $\geq 10\%$ regrowth from nadir & $0.651$ \\
    Volume, $\geq 20\%$ regrowth from nadir & $0.608$ \\
    Volume, $\geq 30\%$ regrowth from nadir & $0.564$ \\
    Volume, $\geq 50\%$ regrowth from nadir & $0.490$ \\
    RECIST diameter, $\geq 20\%$ regrowth   & $0.428$ \\
    \hline
    Range across 14 study/arm codes (20\% criterion) & $0.50$--$0.65$ \\
    \hline
  \end{tabular}
  \label{tab:kather_summary}
\end{table}

\begin{figure}[t]
    \centering
    \includegraphics[width=\textwidth]{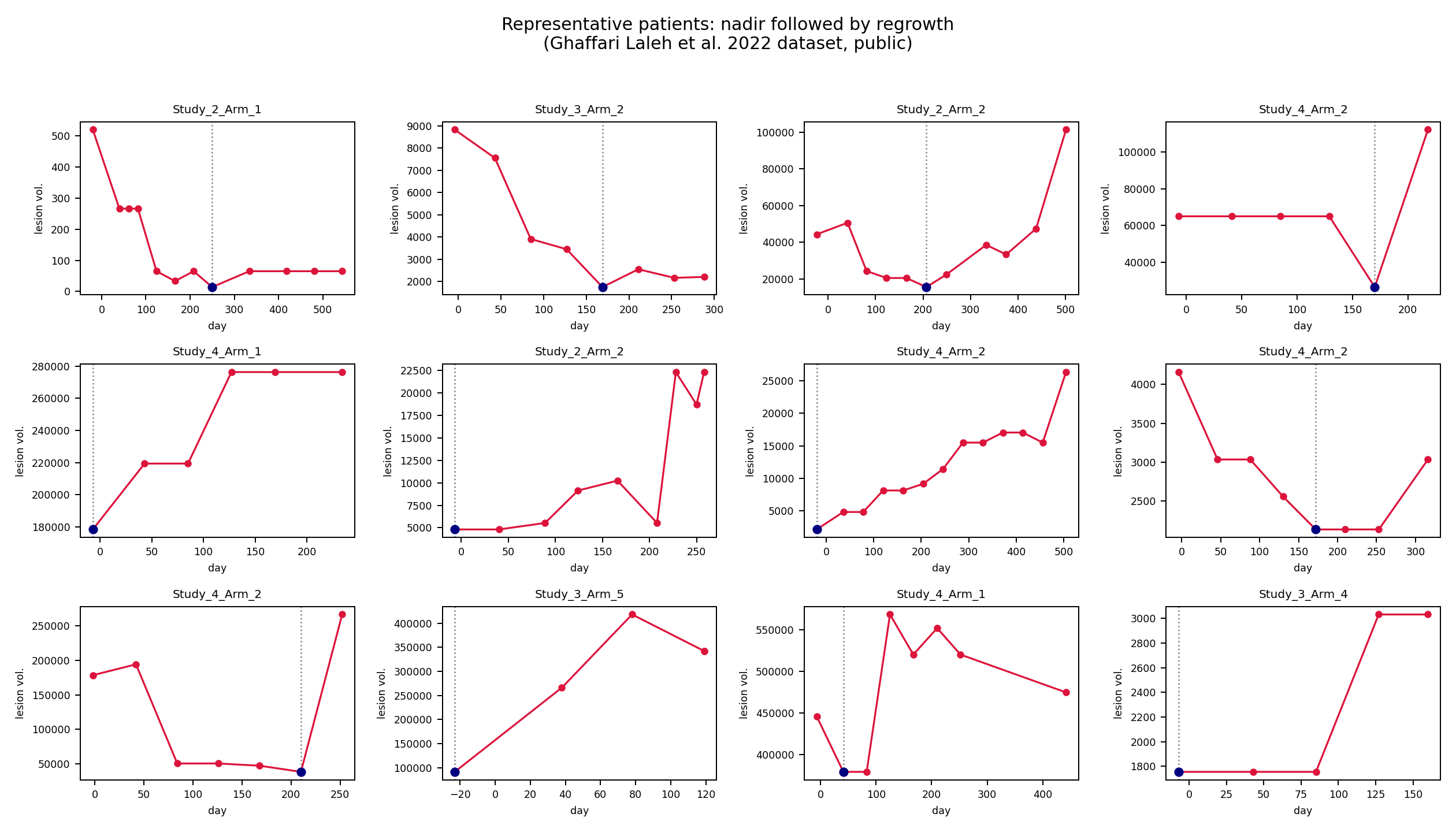}
    \caption{Twelve representative patient trajectories from the public dataset of  \cite{10.1371/journal.pcbi.1009822}, selected among patients classified as showing a nadir-then-regrowth pattern (navy marker: nadir; dotted line: day of nadir). Each panel is a  patient's lesion-volume trajectory under chemotherapy or immunotherapy; study/arm codes are as anonymized in the public release. The qualitative shape — decline to a minimum followed by measurable regrowth before the last recorded visit — is the phenomenological signature that motivates treating $P_{\mathrm{surv}}(t)>0$, rather than $P_{\mathrm{surv}}(t)\to0$, as the generic expectation.}
    \label{fig:kather_examples}
\end{figure}

This analysis  establishes  that the qualitative phenomenon called pharmacokinetic escape  is consistent across independent trial cohorts in real  data. We next go beyond this  descriptive check and attempt an actual quantitative fit of the model's own functional forms.

\subsection{Quantitative confrontation: fitting $\lambda(t)$ and testing the demographic-noise scaling law}
\label{sec:quantitative_fit}

A direct fit of the full survival functional $\Lambda(t)$ (Eq.~\eqref{eq:Lambda_def}) to individual patient trajectories is not well posed with this dataset alone: $\Lambda(t)$ is  related to an ensemble of micro-clusters of a given initial size $C_{\mathrm{ini}}$, not a deterministic prediction for a single observed trajectory, and the dataset provides one realization per patient,
together with no information on the drug dose $D_0$ or absolute micro-cluster scale needed to fix $C_{\mathrm{ini}}$.  But still two  quantitative checks are nevertheless possible to compute.

\subsubsection{Fitting the drift $\lambda(t)$}

For every patient with a nadir-then-regrowth pattern and at least five volume measurements ($n=477$), we fit the model's own bolus drift functional form, Eq.~\eqref{eq:lambda_t} together with Eq.~\eqref{eq:D_exponential}, i.e.\ $\lambda(t) = a_1 - A\,e^{-C_0 t}$ with $A\equiv b_\tau D_0$, directly to the log-volume trajectory via
\begin{equation}
    \log V(t) = \log V(0) + a_1 t - \frac{A}{C_0}\bigl(1-e^{-C_0 t}\bigr) ,
    \label{eq:logV_fit}
\end{equation}
 using nonlinear least squares with $(a_1, A, C_0)$ as free parameters and $V(0)$ fixed to the first observation. This is a purely phenomenological fit of an \emph{effective} net growth rate; it uses no drug-dose information  and does not by itself test the two-compartment PK origin assumed for $D(t)$.

\begin{figure}[t]
    \centering
    \includegraphics[width=\textwidth]{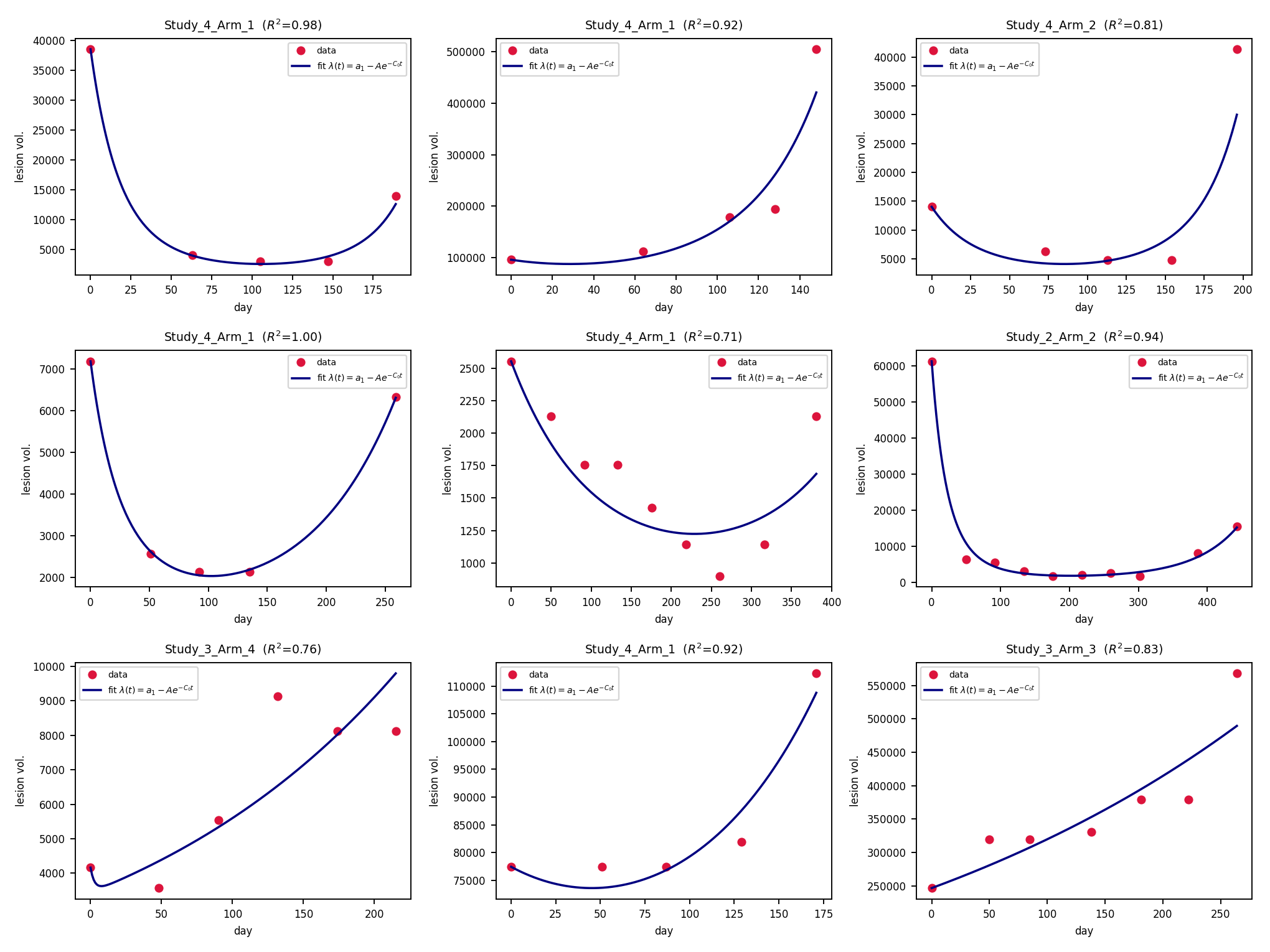}
    \caption{Nine representative fits of the model's bolus drift functional form, Eq.~\eqref{eq:logV_fit}, to individual real patient lesion-volume trajectories from the public dataset of  \cite{10.1371/journal.pcbi.1009822}. Points: observed volumes. Curves: best fit of $\lambda(t)=a_1-Ae^{-C_0t}$ via nonlinear least squares on $\log V(t)$, with $R^2$ noted per panel. The same three-parameter shape assumed for the illustrative bolus example of Sec.~\ref{sec:fokker_planck} reproduces the qualitative decline-then-regrowth pattern across a range of real tumor kinetics.}
    \label{fig:lambda_fits}
\end{figure}

Fits converged for $473/477$ patients; $381$ ($80.5\%$) achieve $R^2 \geq 0.5$, with a median $R^2 \approx 0.87$ among these. Figure~\ref{fig:lambda_fits} shows nine examples: the two-parameter shape of Eq.~\eqref{eq:logV_fit} describes the decline-to-nadir-to-regrowth pattern of real patient trajectories using only three free parameters per patient. Restricting to well-fit patients ($R^2\geq0.5$), $75.6\%$ show a genuine sign change in the fitted drift ($0<a_1<A$, i.e.\ $\lambda(t)$ is negative early and positive late, exactly as assumed for the bolus case in Sec.~\ref{sec:fokker_planck}), with a median crossing time of $93$ days after the first measurement. This is systematically later than the median nadir day of $50$ days found in Sec.~\ref{sec:real_data}; in a purely deterministic reading of Eq.~\eqref{eq:logV_fit} the two should coincide exactly, and we attribute the discrepancy chiefly to the sparse, irregular sampling (a median of five to six visits per patient) limiting the temporal resolution with which the drift zero-crossing can be located, rather than to a qualitative failure of the functional form.

\subsubsection{Testing the demographic-noise}

The paper's central mechanistic claim is that the noise amplitude scales as $\sqrt{\sigma(t)\,C}$ (Eq.~\eqref{eq:CC_cov}, demographic/shot noise), not as a constant fraction of $C$ as in a generic log-normal treatment-response noise model. These two hypotheses make sharply different, falsifiable predictions for how the variance of step-wise log-volume increments should depend on tumor size:
\begin{equation}
    \mathrm{Var}[\Delta \log C] \;\sim\;
    \begin{cases}
        1/C & \text{demographic (shot) noise,} \\
        \text{const.} & \text{constant relative noise.}
    \end{cases}
    \label{eq:noise_scaling_hypotheses}
\end{equation}
For the $381$ well-fit patients of the previous subsection, we compute step-wise residuals of $\log V(t)$ about the \emph{fitted} $\lambda(t)$ curve of Eq.~\eqref{eq:logV_fit} 
, normalize by $\sqrt{\Delta t}$ to combine irregularly spaced visits, pool all step residuals across patients, bin them by the lesion volume at the start of each step, and regress $\log(\text{bin variance})$ against $\log(\text{bin mean volume})$.

\begin{figure}[t]
    \centering
    \includegraphics[width=0.85\linewidth]{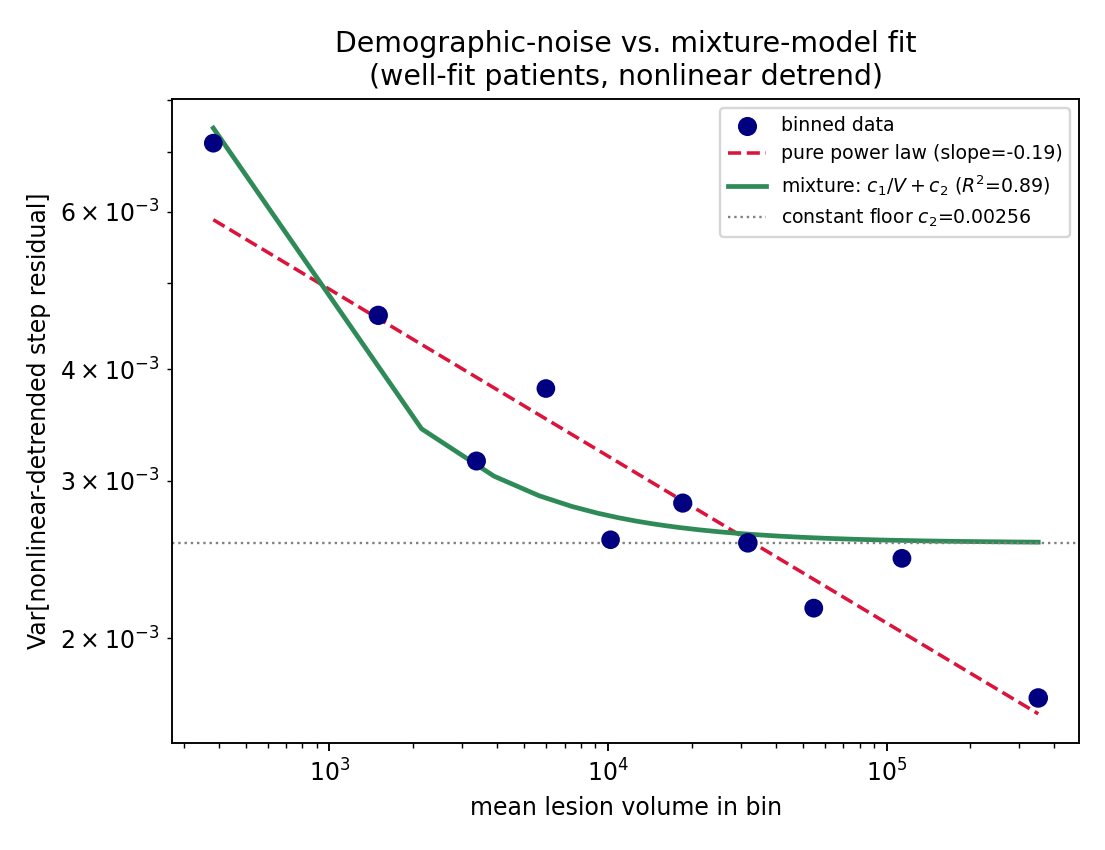}
    \caption{Test of the demographic-noise scaling law, Eq.~\eqref{eq:noise_scaling_hypotheses}, using step-wise residuals about the fitted $\lambda(t)$ curve for the $381$ well-fit patients ($R^2\geq0.5$) of Fig.~\ref{fig:lambda_fits}, binned by lesion volume at the start of each step. The pure power-law fit (slope $-0.19\pm0.02$) is intermediate between the demographic-noise prediction (slope $-1$) and the constant-relative-noise prediction (slope $0$). The two-parameter mixture $c_1/V+c_2$ (Eq.~\eqref{eq:mixture_model}) fits comparably well overall but tracks the data more closely at both small and large volumes, and implies that demographic noise dominates the size-dependent variance at the smallest observed lesion volumes while a roughly size-independent floor dominates at the largest.}
    \label{fig:noise_scaling}
\end{figure}

A pure power-law fit gives a slope of $-0.19 \pm 0.02$ ($R^2=0.88$ for the log-log regression, $n=10$ volume bins), which excludes the constant-noise hypothesis (slope $0$) at high significance ($>8\sigma$) but is also far shallower than the pure demographic-noise exponent of $-1$. Rather than treating this as a simple refutation, we fit the two-component mixture (one given by the demographic noise ($c_1/V$) and the second one is approximately volume-independent noise floor $c_2$ (for instance reflecting segmentation or measurement error in CT-based volumetry))
\begin{equation}
    \mathrm{Var}[\Delta \log C](V) = \frac{c_1}{V} + c_2 ,
    \label{eq:mixture_model}
\end{equation}
This mixture fits the binned data comparably well to the pure power law ($R^2=0.89$ vs.\ $0.88$) but visibly better at both extremes of the observed volume range (Fig.~\ref{fig:noise_scaling}), and yields a directly interpretable decomposition: at the smallest lesion volumes in the well-fit cohort ($\approx 380\,\mathrm{mm}^3$), the demographic term accounts for $\approx 66\%$ of the total predicted variance.

\subsection{Proposed experimental validation}
\label{sec:proposed_validation}

 No existing public dataset provides true replicates under identical conditions that could permit to test our model. In that subsection, we shall propose a dedicated validation study  as a genuine test of the mechanism.

\subsubsection{Tier 1: preclinical, immunocompetent.} 

The most incisive test uses a syngeneic tumor line (not an immunodeficient PDX model, which doesn't have  the NK/T compartment which is  central to our mechanism) implanted in immunocompetent mice with a bioluminescent reporter sensitive to sub-imaging-threshold residual disease. A satellite PK cohort with dense blood sampling would fit the two-compartment model of Eqs.~\eqref{eq:B_dynamics}--\eqref{eq:D_dynamics}, replacing the illustrative $\alpha_a$, $C_0$ of Table~\ref{tab:sim_params} with measured values, while serial non-terminal flow cytometry (NK, CD8) would give $N(t)$, $T(t)$. Table~\ref{tab:validation_arms} summarizes a seven-arm design.  Arm 7  is the control while arms 5--6 isolate the timing-sensitivity of adjuvant immunotherapy.

\begin{table}[t]
  \centering
  \caption{Proposed preclinical validation arms and their falsifiable predictions.}
  \begin{tabular}{p{0.13\linewidth}p{0.4\linewidth}p{0.35\linewidth}}
    \hline
    Arm & Protocol & Model prediction \\
    \hline
    1 & Single bolus (MTD) & baseline relapse, early window \\
    2 & Continuous infusion, $\kappa<\kappa_c$ & relapse comparable to or above arm 1 \\
    3 & Continuous infusion, $\kappa>\kappa_c$ & relapse suppressed near zero \\
    4 & Metronomic & intermediate-to-high relapse \\
    5 & Bolus + \emph{early} adjuvant immunotherapy & strong relapse reduction \\
    6 & Bolus + \emph{late} adjuvant immunotherapy & weak or no reduction \\
    7 & Bolus, immune-depleted/NSG controls & qualitative change in relapse pattern, not only magnitude \\
    \hline
  \end{tabular}
  \label{tab:validation_arms}
\end{table}

An estimation of the number of animals needed for this part of the experiment can be done as follows:

\textbf{Experiment 1: Testing immunotherapy timing.}
The model predicts survival rates of roughly $40\%$ versus $23\%$ depending 
on whether immunotherapy is given early or late. Detecting a difference this 
large with standard statistical thresholds ($\alpha = 0.05$, power $= 0.8$) 
would normally require approximately $102$ animals per group.  Using a multi-site metastasis model, 
where each animal develops roughly $k \approx 5$ spatially distinct, 
measurable tumor foci tracked by bioluminescence can reduce this number. If these foci behave as 
 independent replicates, only $\approx 20$ animals per group will be needed.

\textbf{Experiment 2: measuring $\kappa_c$.}
 A sharp transition at a critical treatment intensity 
$\kappa_c$ is predicted in our model. To locate $\kappa_c$ experimentally, a dose-escalation design 
with $5$--$6$ infusion levels and $n \approx 8$--$10$ animals per level, 
analyzed by logistic regression, is sufficient to estimate $\kappa_c$ with 
a usable confidence interval.  Approximately $50$--$60$ animals 
in total are needed for this step of the experiment.

\textbf{What animal experiments offer beyond human imaging.}
With true biological replicates, the survival probability $P_{\mathrm{surv}}(t)$ 
can be estimated empirically --- simply as the fraction of animals that remain 
alive and below a dangerous tumor size threshold at each time point --- and 
compared  against the theoretical prediction of Eq.~\eqref{eq:P_surv} 
evaluated at the measured parameters $(a_1, b_\tau, C_0)$. Additionally, the 
noise-scaling test can be repeated in parallel arms: immunocompetent animals 
versus immune-depleted animals. If the demographic noise term $c_1$ in 
Eq.~\eqref{eq:mixture_model} shrinks or vanishes under immune depletion, this 
would constitute stronger mechanistic evidence for immune-driven stochasticity.

\subsubsection{Tier 2: Clinical--translational validation.}
A less controlled but more directly generalizable approach embeds additional 
measurements within an existing or planned clinical trial that already compares 
standard maximum-tolerated-dose (MTD) chemotherapy against metronomic 
(continuous low-dose) scheduling~\cite{MetronomicBridge2024, 
MetronomicSystematic2019}. In that case, 
three complementary measurements will be done to the patients participating to this clinical trial.

\textbf{What is measured.}
First, serial circulating tumor DNA (ctDNA) samples are collected from blood 
at regular intervals. Circulating tumor DNA detects residual disease far 
earlier and more sensitively than imaging, which can only identify tumors once 
they are large enough to be visible. Second, serial immune monitoring tracks 
how the patient's immune system is modified during the course of treatment --- for 
example, changes in immune cell populations or activation markers. Third, 
pharmacokinetic (PK) sampling records drug concentrations in the blood as a function of  
time; this is already standard practice in early-phase trials and adds no 
extra burden to patients.

\textbf{How the data are analyzed.}
With these measurements in hand, a hierarchical mixed-effects model is fitted 
separately to each patient's time series. This estimates the patient-specific 
tumor growth rate $\lambda(t)$ and noise level $\sigma(t)$ as functions of 
time, using the actually measured drug exposure and immune activity as inputs. This is analogous to the quantitative 
fitting procedure of Sec.~\ref{sec:quantitative_fit}.

\textbf{What this test achieves.}
The central question is whether the vulnerability window --- the period of 
heightened tumor sensitivity to treatment identified in Tier~1 animal 
experiments --- also appears in human patient data. Finding the 
same pattern in observational clinical data would confirm that the mechanism identified in controlled animal 
models is consistent with how human tumors  behave.

\subsubsection{ Falsification criteria}

 The theory developed 
in this paper would be substantially weakened --- and in some cases 
effectively refuted --- by any of the following four findings.

\textbf{Strike 1: The immune system does not affect relapse patterns.}
 The immune system has a central role in this model. If immune-depleted 
animals (arm~7) show no qualitative difference in relapse pattern compared 
with immunocompetent controls, it will disfavor the model. 

\textbf{Strike 2: The timing of immunotherapy does not matter.}
A central prediction of Sec.~\ref{sec:immunotherapy_numerics} is that there 
exists a vulnerability window. If 
late adjuvant immunotherapy (arm~6) proves as effective as early immunotherapy 
(arm~5), the vulnerability window is not needed
and the timing-sensitivity that motivates the entire scheduling analysis is 
absent from real tumor dynamics.

\textbf{Strike 3: Treatment efficacy increases smoothly with dose, 
with no threshold.}
 A sharp, threshold-like transition in relapse probability 
near a critical treatment intensity $\kappa_c$ is expected. If  relapse probability 
declines continuously  as $\kappa$ increases, with no detectable 
inflection or threshold, it will disfavor the model. 

\textbf{Strike 4: The noise-scaling signal is a measurement artifact.}
The quantitative fitting of Sec.~\ref{sec:quantitative_fit} rests on the 
observation that tumor-to-tumor variability scales with tumor size in a 
specific, predictable way as a signature of genuine biological stochasticity. If the noise-scaling test returns a 
slope statistically indistinguishable from zero even in immunocompetent arms, 
it would indicate that this scaling signal  was predominantly an artifact of 
measurement noise. 

\

\begin{table}[t]
  \centering
  \caption{Parameter values used in the PK--PD SPDE and reduced Feller simulations.}
  \begin{tabular}{llcl}
    \hline
    Symbol & Description & Value & Units \\
    \hline
    $a_1$     & Tumor intrinsic growth rate        & $0.22$   & (time)$^{-1}$ \\
    $a_2$     & Tumor crowding coefficient         & $0.015$  & (cells)$^{-1}$ \\
    $a_3$     & NK-mediated tumor killing          & $0.010$  & (time)$^{-1}$ (cells)$^{-1}$ \\
    $a_4$     & NK intrinsic growth rate           & $0.18$   & (time)$^{-1}$ \\
    $a_5$     & NK crowding coefficient            & $0.020$  & (cells)$^{-1}$ \\
    $a_6$     & Tumor-induced NK depletion         & $0.012$  & (time)$^{-1}$ (cells)$^{-1}$ \\
    $\rho$    & NK decay rate                      & $0.06$   & (time)$^{-1}$ \\
    $\beta_1$ & CTL-mediated tumor killing         & $0.008$  & (time)$^{-1}$ (cells)$^{-1}$ \\
    $\beta_2$ & Tumor-induced CTL depletion        & $0.030$  & (time)$^{-1}$ (cells)$^{-1}$ \\
    $r$       & CTL recruitment from $C N$         & $0.0045$ & (time)$^{-1}$ (cells)$^{-1}$ \\
    $w$       & CTL decay rate                     & $0.05$   & (time)$^{-1}$ \\
    $b_\tau$  & Drug cytotoxicity on tumor         & $0.060$  & (time)$^{-1}$ (drug)$^{-1}$ \\
    $b_\aleph$& Drug cytotoxicity on NK            & $0.050$  & (time)$^{-1}$ (drug)$^{-1}$ \\
    $b_L$     & Drug cytotoxicity on CTL           & $0.050$  & (time)$^{-1}$ (drug)$^{-1}$ \\
    $\alpha_a$& Absorption rate $B\to D$           & $0.60$   & (time)$^{-1}$ \\
    $C_0$     & Active drug decay rate             & $0.18$   & (time)$^{-1}$ \\
    \hline
    $C(0)$    & Initial tumor density              & $18$     & cells \\
    $N(0)$    & Initial NK density                 & $11$     & cells \\
    $T(0)$    & Initial CTL density                & $8$      & cells \\
    \hline
    $\Delta t$& Time step                          & $0.02$   & time \\
    $T_{\text{end}}$ & Final time                  & $120$    & time \\
    \hline
  \end{tabular}
  \label{tab:sim_params}
\end{table}
\begin{figure}[t]
  \centering
  \includegraphics[width=\textwidth]{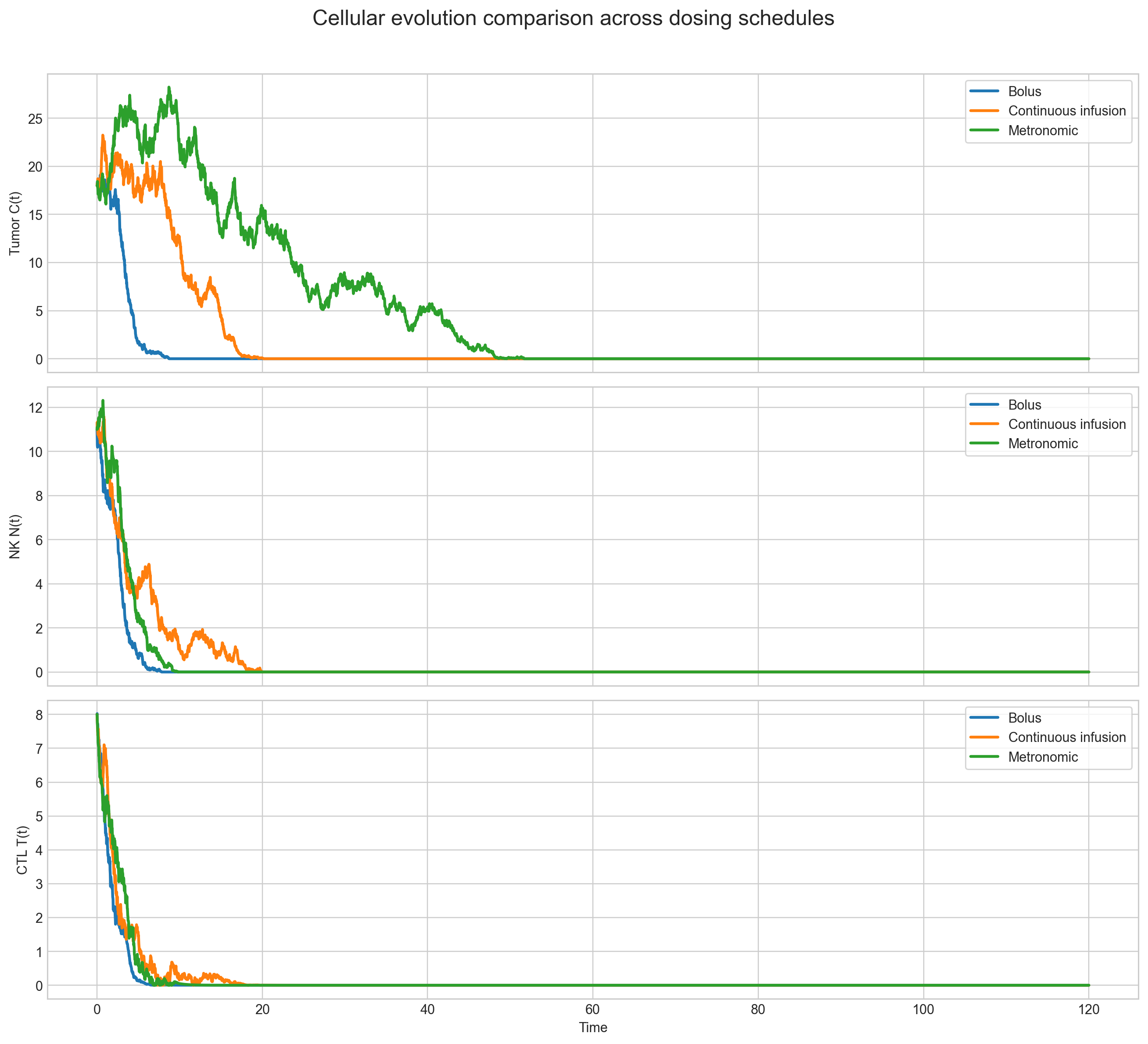}
  \caption{Direct comparison of cellular trajectories across dosing schedules for the same realization of the noise and the parameters in Table~\ref{tab:sim_params}. Top panel: tumor density $C(t)$ under bolus (blue), continuous infusion (orange), and metronomic dosing (green). Middle and bottom panels: NK and CTL dynamics under the same protocols. Bolus dosing produces rapid and strong but transient suppression, continuous
  infusion leads to sustained control, and the metronomic schedule leaves a longer period of subcritical but noisy tumor fluctuations.}
  \label{fig:cellular_comparison}
\end{figure}
\begin{figure}[t]
  \centering
  \includegraphics[width=\textwidth]{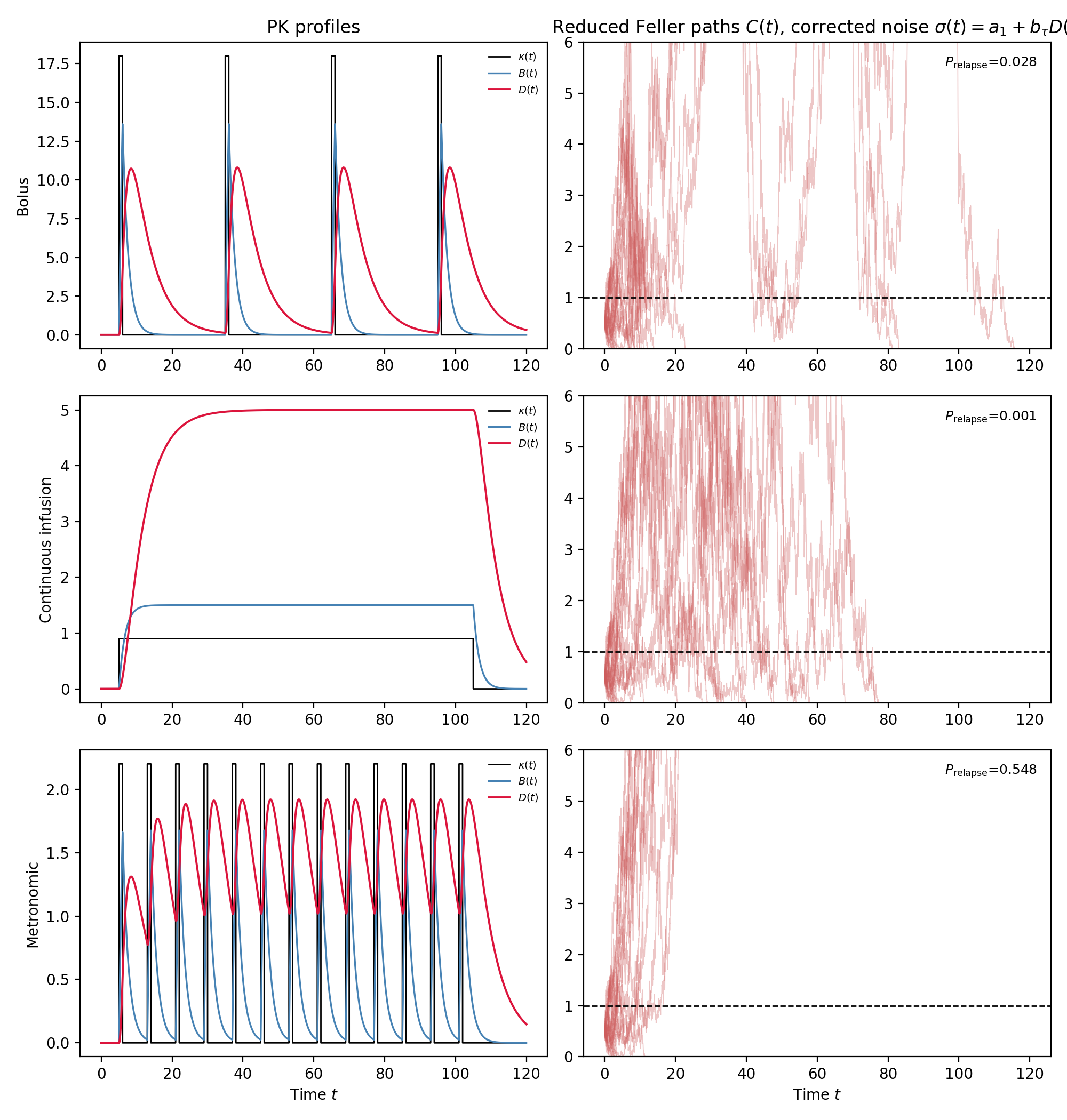}
  \caption{Reduced Feller-description of micro-cluster dynamics in immune-depleted sanctuary sites, using representative bolus, continuous-infusion, and metronomic PK profiles with the parameters of Table~\ref{tab:sim_params}. Left column: pharmacokinetic profiles $\kappa(t)$, $B(t)$, and $D(t)$. Right column: bundles of sample paths ($n=25$ shown; relapse probabilities computed from $n=4000$ trajectories) of the reduced tumor density $C(t)$ governed by $\partial_t C = \lambda(t) C + \sqrt{\sigma(t)\, C}\,\xi(t)$, with $\lambda(t) = a_1 - b_\tau D(t)$ and the corrected noise strength $\sigma(t)=a_1+b_\tau D(t)$ of Eq.~\eqref{eq:sigma_t} (rather than the leading-order $\sigma=a_1$ used in an earlier draft), and an absorbing boundary at $C=0$. The horizontal dashed line marks the relapse threshold at $C=1$, corresponding to the emergence of a detectable micro-cluster. Consistent with the analytical result of Sec.~\ref{sec:fokker_planck}, both bolus and continuous infusion (at $\kappa$ above threshold) strongly suppress relapse, whereas metronomic dosing leaves a substantially higher relapse probability, since the oscillatory drug profile repeatedly re-opens the vulnerability window before the previous pulse's effect has cleared.}
  \label{fig:feller_three_schedules}
\end{figure}
\begin{figure}
    \centering
    \includegraphics[width=1\linewidth]{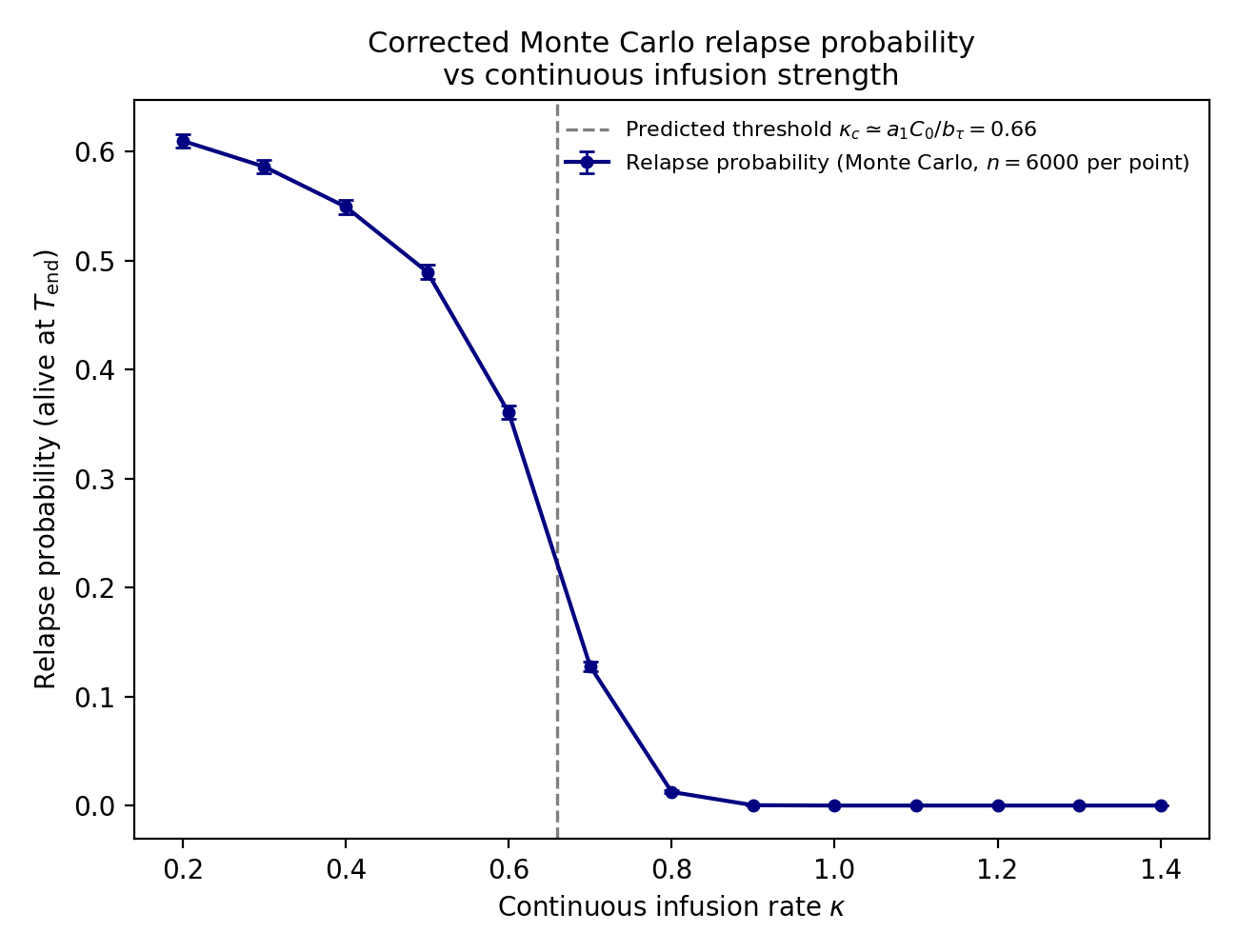}
    \caption{Monte Carlo estimate of the relapse probability for a tumor micro-cluster as a function of the continuous infusion rate $\kappa$ in the reduced Feller description, using the corrected noise strength $\sigma(t)=a_1+b_\tau D(t)$ of Eq.~\eqref{eq:sigma_t}. Each point is estimated from $n=6000$ independent trajectories, with vertical error bars showing the binomial standard error $\sqrt{p(1-p)/n}$; error bars are largest near the transition, where the outcome is most uncertain, and shrink toward both plateaus. The curve shows the fraction of trajectories that both cross the relapse threshold $C=1$ at some time and remain alive at the final observation time $T_{\mathrm{end}}$. As $\kappa$ increases, the relapse probability decreases and becomes negligible once $\kappa$ exceeds the theoretical threshold $\kappa_c \simeq a_1 C_0 / b_\tau$ (vertical dashed line, Eq.~\eqref{eq:kappa_c}), at which the effective steady-state drift changes sign; the location of the Monte Carlo transition is consistent with $\kappa_c$ to within the resolution of the $\kappa$-grid used here.}
    \label{fig:feller_kappa_corrected}
\end{figure}
\section{Conclusions}

We have constructed a non-equilibrium stochastic PK--PD field theory in which pharmacokinetic absorption delay and cellular demographic noise share a common microscopic origin, and used it to derive a closed-form survival functional $P_{\mathrm{surv}}(t)$ (Eq.~\eqref{eq:P_surv}) for tumor micro-clusters in immune-depleted sanctuary sites. As summarized in Sec.~\ref{sec:positioning}, the resulting mechanism — Stochastic Pharmacokinetic Escape — differs from prior stochastic tumor--immune and PK/PD models in coupling the strength of demographic noise  to the dosing protocol via the drug-dependent immune absorbing state. 

Our main results can be summarized as follows: 

\begin{enumerate}

\item  the analytical result that $P_{\mathrm{surv}}(\infty)>0$ for any bolus protocol with finite $\kappa$ and $b_\tau$ (Sec.~\ref{sec:fokker_planck}); 

\item Monte Carlo simulation across bolus, continuous, and metronomic schedules (Fig.~\ref{fig:feller_three_schedules}) and as a function of continuous infusion strength (Fig.~\ref{fig:feller_kappa_corrected}), confirming the threshold $\kappa_c \simeq a_1C_0/b_\tau$ of Eq.~\eqref{eq:kappa_c}; 

\item  direct numerical test of adjuvant immunotherapy (Sec.~\ref{sec:immunotherapy_numerics}, Fig.~\ref{fig:immunotherapy_numerics}), which reduced the relapse probability from $0.405$ to $0.226$ in our reference scenario, but also revealed that its benefit is concentrated in the late relapse window and is limited by how fast immune repopulation can start relative to the pharmacokinetic clearance time $1/C_0$; 

\item  a model-agnostic check against a public dataset of 1461 human patients (Sec.~\ref{sec:real_data}, Fig.~\ref{fig:kather_examples}), showing that $49\%$--$65\%$ of evaluable lesions exhibit a nadir followed by measurable regrowth rather than monotonic elimination, consistently across 14 independent study/arm cohorts; and 

\item  a direct quantitative fit of the model's drift functional form to $473$ real patient trajectories (Sec.~\ref{sec:quantitative_fit}, Fig.~\ref{fig:lambda_fits}), $75.6\%$ of which show the theoretically expected change from net decline to net regrowth, together with a test of the demographic-noise scaling law (Fig.~\ref{fig:noise_scaling}) whose $c_1/V+c_2$ decomposition attributes $\approx 66\%$ of the size-dependent noise variance to a demographic-noise-like term at the smallest observed lesion volumes.
\end{enumerate}

This work illustrates that non-equilibrium statistical field theory can turn a qualitative statement on the fact that demographic noise can prevent  extinction into explicit, protocol-dependent design criteria (Eqs.~\eqref{eq:kappa_c} and \eqref{eq:clinical_gamma_condition}) 
, and that at least some of those criteria can be brought into contact with public clinical data. 

 Natural extensions include explicit spatial heterogeneity in vascularization and immune infiltration, multi-drug combinations with distinct PK profiles, and calibration of $a_1$, $b_\tau$, and the immune-repopulation rates against longitudinal clinical or preclinical data \cite{GeorgeLevine2018,MetronomicReview2014}, which would allow the thresholds derived here to be evaluated for specific drug--tumor cases rather than only illustrative parameter sets. A future direction to validate our approach, detailed in Sec.~\ref{sec:proposed_validation}, would be a dedicated validation study combining directly measured pharmacokinetics, serial immune monitoring, and true biological replicates — most incisively in an immunocompetent preclinical model — to move from the phenomenological and quantitative-but-indirect confrontations of Secs.~\ref{sec:real_data}--\ref{sec:quantitative_fit} to a genuine, falsifiable mechanistic test. We note explicitly that alternative deterministic mechanisms — acquired drug-resistance mutations, clonal selection of pre-existing resistant subpopulations, or immune-checkpoint escape — could also produce nadir-then-regrowth trajectories in the clinical data of Sec.~\ref{sec:real_data}; distinguishing between these and the stochastic noise-driven mechanism proposed here requires the matched pharmacokinetic and immune-monitoring data that the present public dataset does not contain, and that the preclinical Tier~1 design (Table~\ref{tab:validation_arms}) is specifically engineered to provide.
\begin{acknowledgments}
We acknowledge financial support from SECIHTI and SNII (M\'exico). G.D is supported by an FNRS Aspirant (ASP) fellowship (40031451) from the Belgian Fonds de la Recherche Scientifique (FNRS).
\end{acknowledgments}


%

\end{document}